\documentclass[preprint,12pt,authoryear]{elsarticle}

\usepackage{amssymb}
\usepackage{graphicx}%
\usepackage{multirow}%
\usepackage{amsmath,amssymb,amsfonts}%
\usepackage{amsthm}%
\usepackage{mathrsfs}%
\usepackage[title]{appendix}%
\usepackage{xcolor}%
\usepackage{textcomp}%
\usepackage{manyfoot}%
\usepackage{booktabs}%
\usepackage{algorithm}%
\usepackage{algorithmicx}%
\usepackage{algpseudocode}%
\usepackage{listings}%
\usepackage{lscape}
\usepackage{setspace}
\usepackage{lipsum}
\journal{Spatial Statistics}

\begin{document}

\setlength{\tabcolsep}{2pt}
\begin{frontmatter}



\title{Joint spatial modeling of mean and non-homogeneous variance combining semiparametric SAR and GAMLSS models for hedonic prices}

\author[label1]{J.D. Toloza-Delgado}
\affiliation[label1]{organization={Department of Statistics, Faculty of Sciences,  Universidad Nacional de Colombia, Bogota, Colombia},
ead={jdtolozad@unal.edu.co}, ead={https://orcid.org/0000-0001-7523-7625}}

\author[label2]{O.O. Melo}
\affiliation[label2]{organization={Department of Statistics, Faculty of Sciences,  Universidad Nacional de Colombia, Bogota, Colombia},
ead={oomelo@unal.edu.co}, ead={https://orcid.org/0000-0002-0296-4511}}

\author[label3]{N.A. Cruz\corref{cor1}}
\affiliation[label3]{organization={Profesor Visitante, Universitat de les Illes Balears, Departament de Matemàtiques i Informàtica, Phone: +34 637 54 6888, Palma de Mallorca, España},
ead={nelson-alirio.cruz@uib.cat}, ead={https://orcid.org/0000-0002-7370-5111}}

\begin{abstract}
In the context of spatial econometrics, it is very useful to have methodologies that allow modeling the spatial dependence of the observed variables and obtaining more precise predictions of both the mean and the variability of the response variable, something very useful in territorial planning and public policies. This paper proposes a new methodology that jointly models the mean and the variance. Also, it allows to model the spatial dependence of the dependent variable as a function of covariates and to model the semiparametric effects in both models. The algorithms developed are based on generalized additive models that allow the inclusion of non-parametric terms in both the mean and the variance, \textcolor{red}{maintaining the traditional theoretical framework of spatial regression.} The theoretical developments of the estimation of this model are carried out, obtaining desirable statistical properties in the estimators. A simulation study is developed to verify that the proposed method has a remarkable predictive capacity in terms of the mean square error and shows a notable improvement in the estimation of the spatial autoregressive parameter, compared to other traditional methods and some recent developments. The model is also tested on data from the construction of a hedonic price model for the city of Bogotá, highlighting as the main result the ability to model the variability of housing prices, and the wealth in the analysis obtained.
\end{abstract}




\end{frontmatter}

\section{Introduction}\label{sec1}

The spatial regression models popularized by \cite{anselin1988spatial} have allowed to analyze and understand different phenomena and problems in various areas of knowledge such as agriculture \citep{plant2018spatial}, ecology \citep{ver2018spatial} and, specifically, economy, in topics such as regional economic growth \citep{lesage2008spatial}, and unemployment \citep{conley2002socio}. In the study of real estate markets, spatial regression models have been one of the most widely applied techniques in the analysis of housing prices because through the theory of hedonic prices is possible to explain the value of properties according to attributes such as the location, the neighborhood and the internal characteristics (number of floors, number of bathrooms, rooms, among others). Most spatial regression works are based on linear relationships between the exogenous variables and the dependent variable \citep{anselin1988spatial}. Authors such as \citet{clapp2002predicting} introduced non-parametric terms that would depend on the location of the properties; these terms were estimated using Kernel regression.

Recent technological and computational advances have allowed traditional spatial regression methodologies to be expanded and deepened, incorporating Bayesian methodologies \citep{lesage1997bayesian,lesage2015software}, non-parametric and semiparametric models \citep{basile2014modeling,minguez2019alternative}, as well as the specification of models under certain particular conditions that violate some assumptions \citep{kelejian1998generalized}. Regarding heteroscedasticity, some authors such as \cite{sicachapackage} generated packages that allow the estimation of parameters associated with mean and variance in linear spatial regression models. However, this aspect has not been extended to semi-parametric or non-parametric models, as presented in the works of \cite{montero2012sar}, and \cite{basile2018advances}, where constant variance was assumed. \cite{schirripa2023spatial} showed a semiparametric M-quantile spatial model to capture both possible nonlinear effects of the cultural environment on prices and spatial trends in an empirical analysis of market price data, which is robust to heteroscedasticity.
\cite{razen2023multilevel} proposed to analyze not only the expected distribution of real estate market values based on explanatory variables, but also more general distributional characteristics using modern regression models that were beyond the focus on conditional expectations, including the constant of variance parameter. \cite{ONIZUKA2024100793} proposes a Bayesian quantile trend filtering method to estimate the non-stationary trend of quantiles in graphs and apply it to crime data in Tokyo.

It is of standing out that housing prices in each of the projects can have a lot of variability, since common areas are included in the apartments, which can be more attractive and can help raise the price of the property, compared to others nearby. In addition, in some properties in Blackwork, some residents can pay for better finishes than others who cannot, giving this to very different values, between property and property of the same building. However, none of the previous studies propose to model both the mean and the variance and do not allow us to consider the possible spatial autocorrelation that the data has incorporated. In this way, this work allows incorporating the flexibility offered by semiparametric models to integrate the modeling of the average of hedonic prices with the modeling of the variance or standard deviation in spatial regression models, autoregressive spatial regression models (SAR) semiparametric, \textcolor{red}{ where generalized additive model of location, scale, and shape (GAMLSS) are used to estimate all parameters.}

This type of analysis will be of great value in the analysis of real estate markets, where investors could make decisions with data that goes beyond the average price, having a risk measure that would allow them to calculate the variability of the house price, depending on its location and its characteristics such as the area, the number of rooms and the number of bathrooms, among other variables. In addition, it is easier to analyze than quantile regression because it allows variability to be estimated with only two semiparametric regression equations, and not in defined quantiles. Furthermore, the built model is easily adaptable to other types of continuous spatial measurements such as pollution or poverty indices.

The document will be developed in four sections. The first specifies the SAR model, the advances that exist in the modeling of means and variances, the theoretical specification of the semiparametric SAR model with non-homogeneous variance, the theoretical developments that allow the estimation of this new model and the distributional properties of the estimators obtained. Subsequently, a simulation study is carried out to evaluate the capacity of the estimation method, considering the good performance of the estimators. In fourth section, the model is applied in a practical case to understand the behavior of the new house in Bogotá city for 2019; also, the richness of the analysis obtained for the economic analysis of hedonic prices is explored by obtaining estimated effects for mean and variance. Finally, conclusions, recommendations, and future work are presented.
\section{Semiparametric SAR model with non-homogeneous variance}
The SAR model can be expressed in matrix form as:
\begin{equation}
\mathbf{y}= \mathbf{X \boldsymbol{\beta}}+\rho \mathbf{Wy}+\mathbf{u}
\label{EC2}
\end{equation}
where $\mathbf{u} \sim N(\mathbf{0},\sigma^2\mathbf{I})$, $\boldsymbol{\beta}$ is a parameters vector of of size $k \times 1$ that has $k-1$ explanatory variables, $\mathbf{X}$ is a matrix with the explanatory variables, $\rho$ is a autoregressive parameter and $\textbf{W}$ is a matrix of spatial weights. \textcolor{red}{In \cite{banerjee2003hierarchical}, the estimation conditions for these models are discussed.}.
The parameter estimation can be made by maximum likelihood, generalized method of moments (GMM) or minimal squares in two stages \citep{anselin1988spatial}, taking into account that the estimate for ordinary least squares is inconsistent because the term $\mathbf{Wy}$ generates endogeneity problems.

For the case of maximum likelihood, the estimation procedure is summarized in four steps according to \cite{arbia2006spatial}, where a partial or concentrated log-likelihood function must be maximized to get the estimate of parameter $\rho$.

This model was extended by \cite{montero2012sar} and \cite{minguez2019alternative}, allowing the incorporation of non-parametric terms that depend on up to three variables (length, latitude, and time in some cases). In this way, the predictive capacity concerning its linear counterpart is improved, and spatial heterogeneity is taken into account. This model was used in hedonic price analysis in Madrid \citep{montero2018housing} and recently in Bogotá \citep{delgado2021determinants}. However, the current limitation of this model lies in the fact that it assumes that variance is constant, something in practice, it is unusual. Regarding, the joint modeling of mean and variance, several techniques allow us to address this problem, among which are the GAMLSS \citep{rigby2005generalized} models, the interconnected generalized linear models (GLM)  \citep{lee2017data} and double-generalized linear models \citep{cepeda2012double}. \textcolor{red}{These models easily allow the inclusion of non-parametric components, but the spatial component is modeled through random effects with conditional autoregressive (CAR), intrinsic autoregressive (IAR), Markov random field (MRF), type structures or, through Gaussian Markov Random Field, where the latter is completely incorporated into the GAMLSS \citep{de2018gaussian}. This is mainly because the conditional distributions for SAR-type random effects do not have a convenient shape, and this prevents them from being easily extended to other distributions beyond the normal one and to GLM \citep{banerjee2003hierarchical}. This is reinforced by what was stated by \citet{haining2020modelling}, who made a comparison of the hierarchical models and the approximation from spatial econometrics, finding equivalences between some structures such as the spatial error model (SEM) or spatial lag of X (SLX) models. However, in the SAR model, this equivalence does not exist because hierarchical modeling separates the data model (the likelihood) and the data-generating process (the process model) into two conditional probability models, where spatial dependence is modeled in the process model through covariates and structured random effects. In contrast, from the point of view of spatial econometrics, this separation is not made and the process model is integrated into the data model, where the values of the results have to be modeled jointly.}

In this context, the work carried out by \cite{cepeda2012double} proposed a methodology to model average and variance in spatial regression models, based on doubly generalized models and Bayesian estimation. However, again under linear relations between explanatory and dependent variables are highlighted. In this way, the absence of a spatial auto-regressive model that includes the two components that have been mentioned previously is highlighted: non-parametric functions of explanatory variables and the joint modeling of mean and variance. Therefore, a semiparametric SAR model with non-homogeneous variance is proposed.

The starting point is semiparametric regression with substantive correlation and heteroscedastic variance, which is given by the following expression:
\begin{align}
 &\mathbf{y}=\rho\mathbf{Wy}+\mathbf{X\boldsymbol\beta}+\sum_{j=1}^p{f}_j(\mathbf{x}_j)+\boldsymbol{\omega}, \quad \boldsymbol{\omega}\sim N(\mathbf{0},\mathbf{\Sigma})
\label{ECGSAR1} \\
&(\mathbf{I}-\rho\mathbf{W})\mathbf{y}=
\mathbf{A} \mathbf{y}=\mathbf{X\boldsymbol\beta}+\sum_{j=1}^p{f}_j(\mathbf{x}_j)+\boldsymbol{\omega}, \qquad \mathbf{A}= \mathbf{I}-\rho\mathbf{W}\nonumber\\
&\mathbf{\Sigma}^{-\frac{1}{2}}\left(\mathbf{A} \mathbf{y}-\mathbf{X\boldsymbol\beta}-\sum_{j=1}^p{f}_j(\mathbf{x}_j)\right) =
\mathbf{\Sigma}^{-\frac{1}{2}}\boldsymbol{\omega} =  \mathbf{v} \label{EC4}
\end{align}
where $\mathbf{X}$, $\boldsymbol{\beta}$ and $\mathbf{W}$  are analogous to those defined in Equation \eqref{EC2}, $\mathbf{\Sigma}$ a diagonal matrix with elements $\Sigma_{ii}=\sigma_i^2=\exp({\boldsymbol{\alpha}^\top \mathbf{x}_{\sigma_i} }+\sum_{j=1}^q{g}_j(\mathbf{x}_{\sigma_{ij}}))$, where the subscript $\sigma$ in expressions like $\mathbf{x}_{\sigma_i}$ indicates that this matrix contains the explanatory variables associated with the variance, $f_j(x)$ and $g_j(x)$ are arbitrary functions and $|\rho|<1$.

This model has an additional challenge concerning the proposal of \cite{montero2012sar} because it incorporates the estimation of the parameters associated with variance, which can be both linear parametric or non-parametric and more flexible.  Next, the likelihood function is obtained, starting from the fact that the variance of the error is given by $\text{E}[\boldsymbol{\omega}\boldsymbol{\omega}^\top]\mathbf{\Sigma}$.
Following \cite{anselin1988spatial}, although $\mathbf{v
}$ is a vector of independent errors with a standard normal distribution, these cannot be observed and the likelihood function will have to be based on $\mathbf{y}$. For this reason, the Jacobian is introduced, which allows the joint distribution of $\mathbf{y}$ to be derived from $\mathbf{v}$. Thus, using the equation \eqref{EC4}, it is obtained that:
\begin{equation}
    \vert J\vert=\left|\frac{\partial\mathbf{v}}{\partial\mathbf{y}}\right|=|\mathbf{\Sigma}^{-1/2}\mathbf{A}|=|\mathbf{\Sigma}^{-1/2}||\mathbf{A}|=|\mathbf{\Sigma}|^{-1/2}|\mathbf{A}|
    \label{EC4}
\end{equation}
Consequently, based on the standard normal distribution of the error term $\mathbf{v}$, and using the result of the equation \eqref{EC4}, the log-likelihood function for the observations vector $\mathbf{y}$ is:
$$\ell=-\frac{n}{2}\ln(\pi)-\frac{1}{2}\ln|\mathbf{\Sigma}|+\ln|\mathbf{A}|-\frac{1}{2}\mathbf{v}^\top\mathbf{v}$$
Therefore, by introducing the penalization and reexpressing non-parametric terms as a líneal combination of splines base functions, the above equation can be rewritten as:
\begin{align*}
    \ell_p&=-\frac{n}{2}\ln(\pi)-\frac{1}{2}\ln|\mathbf{\Sigma}|+\ln|\mathbf{A}|-\frac{1}{2}\mathbf{v}^{\top}\mathbf{v}-\\
    &\frac{1}{2}\left(\boldsymbol{\beta}^{\top}\left[\sum_{j=1}^p\psi_{1j}\mathbf{G}_{1j}\right]\boldsymbol{\beta}+\sum_{j=1}^q \boldsymbol{\alpha}_j^{T}\psi_{2j}\mathbf{G}_{2j}\boldsymbol{\alpha}_j\right)
\end{align*}
with  $\mathbf{v}^{\top}\mathbf{v}=(\mathbf{Ay}-\mathbf{\tilde{X}\boldsymbol{\beta}})^\top\mathbf{\Sigma^{-1}}(\mathbf{Ay}-\mathbf{\tilde{X}\boldsymbol{\beta}})$ and where $\mathbf{\tilde{X}}$ is a matrix that includes variables whose relation is assumed to be linear in the model of the mean and those that were reparameterized by base functions associated with non-parametric terms, $\pmb{\beta}$ is the vector of parameters associated with the explanatory variables for the mean, $\pmb{\alpha}_j$ is the vector of parameters associated with the explanatory variables for the variance,  $\psi_{1j}$ is the smoothing parameters and $\mathbf{G}_{1j}$  is a matrix associated with the penalty of P-splines, defined following the idea of \citet{durban2015metodos} for $\pmb{\beta}$, and analogously, $\psi_{1j}$ and $\mathbf{G}_{1j}$  are associated with the penalty for $\pmb{\alpha}_j$. In the case of the matrix associated with variance, it is also reparametrized with the base functions and will be worked as $\Sigma_{ii}=\sigma_i^2=\exp({\boldsymbol{\alpha}^\top \tilde{\mathbf{x}}_{\sigma_i} })$.

Next, the respective predictors are proposed for each of the parameters of interest ($\mu_i$ and $\sigma_i^2$), according to the GAMLSS models:
\begin{align*}
    \mathbf{y}&\sim  \mathcal{D}(\boldsymbol{\mu},\boldsymbol{\sigma}^2)\\
\boldsymbol{\eta}_1 & =g_1(\boldsymbol{\mu})=\boldsymbol{\mu}=\rho\mathbf{Wy}+\mathbf{\tilde{X}\boldsymbol{\beta}}\\
\boldsymbol{\eta}_2 & =g_2(\boldsymbol{\sigma})=\mathbf{\tilde{X}_\sigma\boldsymbol{\alpha}}
\end{align*}
where the auto-regressive spatial parameter ($\rho$) is incorporated into the mean. To maximize this function, it must be performed using numeric methods (mainly the auto-regressive parameter $\rho$). In addition, the fact of having non-parametric terms incorporates the estimation of the comfort parameters, which can be obtained from cross-validation, generalized cross-validation, restricted maximum likelihood (REML), Laplace approximations, or the minimization of the generalized criterion of information of Akaike \citep{stasinopoulos2017flexible}. The method proposed in this work corresponds to an adaptation of the algorithm proposed by \cite{anselin1988spatial} to estimate SAR models, \textcolor{red}{which uses GAMLSS models to estimate the parameters associated with the mean and variance} \citep{rigby2005generalized}. The methodology is based on the fact that by knowing $\rho$ and multiplying the vector $\mathbf{y}$ in \eqref{EC2} by the matrix $\mathbf{A}=(\mathbf{I}-\rho\mathbf{W})$, the model parameters can be obtained through a model for the mean and variance.
Thus, by deriving the log-likelihood function for the parameters of interest, the following equations are obtained:
\begin{align*}
\frac{\partial{\ell_p}}{\partial\boldsymbol{\beta}}&=\tilde{\mathbf{X}}^T\mathbf{\Sigma}^{-1}\mathbf{v}-\sum_{j=1}^p\psi_{1j}\mathbf{G}_{1j}\boldsymbol{\beta}\\
\frac{\partial{\ell_p}}{\partial\rho}&=-\operatorname{tr} (\mathbf{A}^{-1} \mathbf{W})+\mathbf{v}^{\top} \mathbf{\Sigma}^{-1 / 2} \mathbf{W} \mathbf{y}\\
\frac{\partial{\ell_p}}{\partial\boldsymbol{\alpha}_m}&=-(1 / 2) \operatorname{tr} (\mathbf{\Sigma}^{-1} \mathbf{H}_{{m}})+(1 / 2) \mathbf{v}^{\top} \mathbf{\Sigma}^{-3 / 2} \mathbf{H}_{{m}}(\mathbf{Ay}-\tilde{\mathbf{X}}\boldsymbol{\beta})-\frac{1}{2}\psi_{2m}\mathbf{G}_{2m}\boldsymbol{\alpha}_m
\end{align*}
for $m=1,2,...,q$, where $\mathbf{H}_m$ is a diagonal matrix with elements $\frac{\partial\exp(\boldsymbol{\alpha}^{\top} \tilde{\mathbf{x}}_{\sigma_i})}{\partial \alpha_m}$, where this last equation is equivalent to the score function obtained by \cite{aitkin1987modelling} when differentiating with respect to the vector of parameters associated with the variance.
The above system is nonlinear, mainly in the autoregressive parameter $\rho$. Additionally, the estimation of the vector of $\boldsymbol{\alpha}$ is complicated since it depends on the $\boldsymbol{\beta}$, and these in turn depend on $\rho$. However, even though the system of equations is nonlinear, the vector of $\boldsymbol{\beta}$ has a closed solution and is given by:

$$\hat{\boldsymbol{\beta}}=\left(\tilde{\mathbf{X}}^{\top}\mathbf{\Sigma}^{-1}\tilde{\mathbf{X}}+\sum_{j=1}^p\psi_{1j}\mathbf{G}_{1j}\right)^{-1}\tilde{\mathbf{X}}^{\top}\mathbf{\Sigma}^{-1}\mathbf{Ay}$$
In this way, the following iterative algorithm is proposed based on the joint estimation of mean and variance modeled by GAMLSS:
\begin{enumerate}
    \item Estimate a GAMLSS for mean and variance between the dependent variable $\mathbf{y}$ and the explanatory variables ($\mathbf{\tilde{X}}$ and $\mathbf{\tilde{X}}_\sigma $), which contains linear and non-parametric terms. Estimate the smoothing parameters $\boldsymbol{\psi}_1$ and $\boldsymbol{\psi}_2$ as well as the variances $\hat{\sigma}_i^2$ to obtain $\hat{\mathbf{\Sigma }}$.
   \item Construct the joint log-likelihood function $(\ell_c)$ and replace $\mathbf{\Sigma}$ with the matrix $\hat{\mathbf{\Sigma}}$ obtained in the previous step and the parameters of smoothing. The joint log-likelihood function is given by:
$$\ell_c=-\frac{n}{2}\ln(\pi)-\frac{1}{2}\ln|\mathbf{\Sigma}|+\ln|\mathbf{A}|- \frac{1}{2}\mathbf{v}^{\top}\mathbf{v}$$
where $$\mathbf{v}^{\top}\mathbf{v}=(\mathbf{Ay}-\mathbf{\tilde{X}}\boldsymbol{\beta})^{\top}\mathbf{\Sigma}^{-1}(\mathbf{Ay}-\mathbf{\tilde{X}}\boldsymbol{\beta})$$ $$\hat{\boldsymbol{\beta}}=\left(\mathbf{ \tilde{X}}^{\top}\mathbf{\Sigma}^{-1}\mathbf{\tilde{X}}+\sum_{j=1}^p\psi_{1j}\mathbf{G}_{1j}\right) ^{-1}\mathbf{\tilde{X}}^{\top}\mathbf{\Sigma}^{-1}\mathbf{Ay}$$

   \item Maximize the joint log-likelihood function and find $\hat{\rho}$. $\hat{\rho}\approx\hat{\rho}_f$, until convergence occurs.
   \item With the $\hat{\rho}$ found in the previous step, the matrix $\mathbf{A}=(\mathbf{I}-\rho\mathbf{W})$ is created and multiplied by $\mathbf{y}$ to obtain $\mathbf{Ay}$.
    \item Taking into account that $\mathbf{Ay}\sim N(\mathbf{\tilde{X}}\boldsymbol{\beta},\mathbf{\Sigma})$ when estimating a GAMLSS model for mean and variance, it will obtain smoothing estimators and a more precise $\boldsymbol{\alpha}$ vector, since the matrix $\mathbf{A}$ acts as a filter that removes the effects of spatial correlation and allows the parameters to be obtained directly of interest. In this way, $\boldsymbol{\psi}_1$ and $\boldsymbol{\psi}_2$ are estimated again as well as the variances $\hat{\sigma}_i^2$ to obtain $\hat{\mathbf{ \Sigma}}$.
    \item With the smoothing parameters and the matrix $\hat{\mathbf{\Sigma}}$, the joint log-likelihood function is constructed again, and $\hat{\rho}_f$ is estimated.

 \item $\hat{\rho}_f$ and $\hat{\rho}$ are compared and steps 5 to 7 are iterated until $\hat{\rho}_f \approx \hat{\rho}$.
 \item  Given $\hat{\rho}_f$, estimate a GAMLSS model with the dependent variable given by $(\mathbf{I}-\rho \mathbf{W})\mathbf{y}$ and obtain the estimators $ \boldsymbol{\hat{\beta}}$ and $\boldsymbol{\hat{\alpha}}$ for the mean and variance, respectively.
\end{enumerate}
The proposed algorithm allows all parameters to be estimated jointly in an iterative process, which provides better estimates of $\rho$. However, for large volumes of information, the calculation of $\ln|\mathbf{\Sigma}|$ generates certain computational problems, since by directly carrying out the operation the software approaches it to infinity, preventing the log-likelihood function from being maximized. Therefore, using the Cholesky decomposition this expression can be obtained with:
$$\ln|\mathbf{\Sigma}|=2\sum\ln\text{diag}(\mathbf{L})$$
where $\mathbf{L}$ is a triangular matrix and $\mathbf{L}^\top \mathbf{L}=\mathbf{\Sigma}$.
The inference of the model is based on the results of \cite{wood2017generalized} and \cite{stasinopoulos2017flexible}, regarding confidence intervals and hypothesis tests on parameters and non-parametric components, with the difference that there is now, an additional parameter ($\hat{\rho}$). These results are obtained through the inverse of the Fisher information matrix $(\mathbf{\mathcal{I}}^{-1})$.

Based on the approach of \cite{anselin1988spatial}, adapting for the semiparametric SAR model with non-homogeneous variance, the following components of the matrix $\mathbf{\mathcal{I}}$ are obtained:
\begin{align*}
\mathcal{I}_{\boldsymbol{\beta} \boldsymbol{\beta}^{\top}}&=\mathbf{\tilde{X}}^{\top}\mathbf{\Sigma}^{-1}\mathbf{\tilde{X}}+\boldsymbol{\psi}_1\mathbf{G}_1\\
\mathcal{I}_{\boldsymbol{\beta} \rho}&= \mathbf{\tilde{X}}^{\top} \mathbf{\Sigma}^{-1} \mathbf{W}\mathbf{A}^{-1} \mathbf{\tilde{X}} \boldsymbol{\beta}\\
\mathcal{I}_{\boldsymbol{\beta} \boldsymbol{\alpha}^{\top}}&=\mathbf{0}\\
\mathcal{I}_{{\rho}{\rho}}&=\text{tr}(\mathbf{W}\mathbf{A}^{-1})^2+\textbf{tr}(\mathbf{\Sigma}(\mathbf{W}\mathbf{A}^{-1})^\top\mathbf{\Sigma}^{-1}(\mathbf{W}\mathbf{A}^{-1}))+(\mathbf{W}\mathbf{A}^{-1}\mathbf{\tilde{X}}\boldsymbol{\beta})^\top\mathbf{\Sigma}^{-1}(\mathbf{W}\mathbf{A}^{-1}\mathbf{\tilde{X}}\boldsymbol{\beta})\\
\mathcal{I}_{{\rho}{\boldsymbol{\alpha}_m}}&=\text{tr}(\mathbf{\Sigma}^{-1}\mathbf{H}_m\mathbf{W}\mathbf{A}^{-1})\\
\mathcal{I}_{{\boldsymbol{\alpha}_m}{\boldsymbol{\alpha}_o}}&=\frac{1}{2}(\text{tr}(\mathbf{\Sigma}^{-2}\mathbf{H}_m\mathbf{H}_o)+\boldsymbol{\psi}_2\mathbf{G}_2)
\end{align*}
where the last equation can be reexpressed as:
$$\mathcal{I}_{{\boldsymbol{\alpha}_m}{\boldsymbol{\alpha}_o}}=\frac{1}{2}\left(\mathbf{\tilde{X}}^\top_{\sigma}\mathbf{\tilde{X}}_{\sigma}+\boldsymbol{\psi}_2\mathbf{G}_2\right)$$
In this way, by inverting the information matrix, the variance and covariance matrix of the estimators $(\mathbf{V}_{\hat{\boldsymbol{\beta}}})$ can be obtained, which allows generating the confidence intervals and perform the relevant hypothesis tests, both for the parameters associated with the mean and the variance or standard deviation.
Once the covariance matrix of the estimators is obtained, it is possible to calculate the standard errors associated with the parameters $\boldsymbol{\beta}$ and $\boldsymbol{\alpha}$ that are linearly incorporated into the model. In the case of non-parametric curves or functions, the estimated function is:
$$\hat{\mathbf{f}}=\acute{\mathbf{X}}\hat{\boldsymbol{\beta}}$$
where $\acute{\mathbf{X}}$ is a matrix that has zeros in the columns corresponding to the coefficients that do not intervene in the estimation of $f(x)$, while the other columns contain the generated base functions to estimate $f$. The standard errors for the estimated curves are given by:
$$\mathbf{v}^*=diag(\acute{\mathbf{X}}\mathbf{V}_{\hat{\boldsymbol{\beta}}} \acute{\mathbf{X}}^T)$$
with these values, it is possible to generate the confidence intervals for the estimated functions, which will be:
$$\hat{f}_i\pm z_{\alpha/2}\sqrt{v_i^*}$$
with $v_i^*$ the $i$th element of the diagonal of the matrix $\mathbf{v}^*$ and $ z_{\alpha/2}$ being the $\alpha/2$ quantile of the standard normal distribution. Although only the $\boldsymbol{\beta}$ parameters are mentioned, the procedure is the same for the $\boldsymbol{\alpha}$ associated with the variance or standard deviation.
Now, to perform a hypothesis test, the following results are obtained:
\begin{itemize}
    \item[i.] On the parameter $\rho$, given that $\mathbf{V}_{\hat{\boldsymbol{\beta}}}$ could be derived from the information matrix, the following Wald test can be performed:

$$z=\frac{\hat{\rho}-\rho_0}{se(\hat{\rho})}$$

where $se(\hat{\rho})$ is the standard error of the spatial autoregressive parameter. Another way to determine the statistical significance of the parameter ($\rho$) is to use the likelihood ratio test, which can be expressed as the difference between the global deviances of the model under the null hypothesis ($H_0:\, \rho=0$) and the alternative hypothesis ($H_0:\, \rho\neq 0$). In this way, the proof is expressed as:
\begin{align*}
    \Lambda&=GD_0-GD_1, \qquad GD=-2\ell(\boldsymbol{\hat{\theta}}),\\
    \ell(\hat{\boldsymbol{\theta}})&=\Sigma_{i=1}^{n}
    \ell\left(\hat{\boldsymbol{\theta}}^{i}\right),  \qquad\hat{\boldsymbol{\theta}}=(\boldsymbol{\mu},\boldsymbol{\sigma})
\end{align*}
where:
\begin{align*}
    GD_0 & =-2\left[-\frac{n}{2}\ln(\pi)-\frac{1}{2}\ln|\mathbf{\Sigma}|-\frac{1}{2}(\mathbf{y}-\mathbf{\tilde{X}}\boldsymbol{\beta})^\top\mathbf{\Sigma^{-1}}(\mathbf{y}-\mathbf{\tilde{X}}\boldsymbol{\beta})\right]\\
    & -2\left[ -\frac{1}{2}\left(\sum_j^p\psi_{1j}\boldsymbol{\beta}_{j}\mathbf{G}_{1j}\boldsymbol{\beta}_{j}+\sum_j^q\psi_{2j} \boldsymbol{\alpha}_{j}\mathbf{G}_{2j}\boldsymbol{\alpha}_{j}\right)\right],\\
    GD_1 & =-2\Bigg[-\frac{n}{2}\ln(\pi)-\frac{1}{2}\ln|\mathbf{\Sigma}|+\ln|\mathbf{A}|-\frac{1}{2}\mathbf{v}^\top\mathbf{v}\\
    &-\frac{1}{2}\left(\sum_j^p\psi_{1j}\boldsymbol{\beta}_{j}\mathbf{G}_{1j}\boldsymbol{\beta}_{j}+\sum_j^q\psi_{2j}\boldsymbol{\alpha}_{j}\mathbf{G}_{2j}\boldsymbol{\alpha}_{j}\right)\Bigg]
\end{align*}
Now, following \cite{rigby2005generalized}, the likelihood ratio statistic follows a chi-square distribution, $\Lambda\sim \chi^2$, where the degrees of freedom are given by the difference between the degrees of freedom of the model under $H_0$ and $H_1$.
\item[ii.] To test the hypothesis that $H_0: \mathbf{C\boldsymbol{\beta}}_j=\mathbf{d}$, where $\boldsymbol{\beta}_j$ is a subvector of $\boldsymbol{\beta}$ which contains only the non-penalized parameters (fixed effects) \citep{wood2017generalized}. In this way, a hypothesis test is derived for $F$ and it is given by:
$$(\mathbf{C}\boldsymbol{\hat{\beta}}_{j}-\mathbf{d})^{\top} (\mathbf{C}\mathbf{V}_{\hat{\boldsymbol{\beta}_j}}\mathbf{C}^{\top})^{-1}(\mathbf{C}\boldsymbol{\hat{\beta}}_{j}-\mathbf{d})$$
just as in the case of the general linear model. Likewise, when statistical hypothesis testing of individual significance is needed, it is possible to use traditional tests based on the standard normal (Wald test) and also perform likelihood ratio tests.
\item[iii.] To perform hypothesis tests on non-parametric components in nested models, \cite{hastie1990generalized}, \cite{rigby2005generalized} and \cite{wood2017generalized} proposed to perform likelihood ratio tests based on the comparison between the global deviance of the maximal model ($\hat{D}_1$) and the minimal ($\hat{D}_0$), taking care with the degrees of freedom effective of the models ($df_{1}^{\mathrm{err}}$ and $df_{0}^{\mathrm{err}}$). In this way, a model for mean and variance will have the following degrees of freedom \citep{rigby1996semi}:
$$d f^{\mathrm{err}}=n-d f_{\mu}-d f_{\sigma^{2}}$$
where $d f_{\mu}$ and $d f_{\sigma^{2}}$ are the effective degrees of freedom used in each of the submodels, respectively, which are given by:
\begin{align*}
d f_{\mu} &=d_{1}+\sum_{j=1}^{p} d f_{1 j} \\
d f_{\sigma^{2}} &=d_{2}+\sum_{k=1}^{q} d f_{2 k}
\end{align*}
where $d_1$ and $d_2$ are the degrees of freedom used in the parametric components of the submodels, where $d f_{1 j}$ and $d f_{2 k}$ are the degrees of freedom used for the smoothed functions.
Consequently, the likelihood ratio test is given by:
$$\Lambda=\hat{D}_{0}-\hat{D}_{1}$$
Thus, following \cite{hastie1990generalized}, the likelihood ratio test can be used as a model selection tool.
\end{itemize}
In next section, a simulation study is carried out to evaluate the estimates of the proposed model.
\section{Simulation Study}
\textcolor{red}{The parameters were estimated using the methodology previously constructed for each sample. The mean of the estimator, the standard deviation of the estimator, and the mean bias were calculated. In addition to evaluating performance against the original simulation parameters, the following models will be adjusted: i) AM-SAR: additive model with spatially lagged dependent variable \citep{montero2012sar}, ii) ML: maximum likelihood SAR model \citep{anselin1988spatial}, iii) H-AM-SAR: semiparametric autoregressive model with heteroscedasticity (methodology proposed in this paper), and iv) GAMLSS: GAMLSS model with $\mathbf{Wy}$ as explanatory variable \citep{rigby2005generalized}.
The simulation will be carried out in three spatial scenarios: i) on a regular grid, ii) on an irregular grid (The results are shown in the appendix \ref{simCundi}), and iii) on spatial points on a given map.
}
\subsection{Analysis on a regular grid}
Each $y_i$ generated from a normal distribution on a regular grid was simulated, following the steps developed by \cite{goulard2017predictions}.
$$y_i\sim N\left(\mu_i=\rho \sum _{j=1}^n{w_{ij}y_i} + \beta_0-\beta_1 x_{1i}+\beta_2x_{2i} +f(x_{3i});\sigma_i=\exp(\alpha_0-\alpha_1x_{2i})\right)$$
where $x_{1i}\sim N(0,1)$, $x_{2i}\sim N(2,1) $, $x_{3i}\sim U(0,1)$ and the $w_{ij}$ follow a tower-like first-order contiguity.
\textcolor{red}{$i=1, \ldots, n$, $n=81, 144, 225, 400$, each value of $\rho= -0.8, -0.4, -0.2, 0.2, 0.4, 0.8$, $\beta_0=2$, $\beta_1=0.5$, $\beta_2=1.75$, $\alpha_0=0.5$ and $\alpha_1=0.3$. 500 simulations of each combination of parameters were run. The variable $x_{3i}$ is a variable that affects the response variable in a non-parametric way by means of the function $f(x_{3i})=0.2x_{3i}^{11}(10(1 - x_{3i}))^6 + 10(10x_{3i})^ 3(1 - x_{3i})^{10}$. These results are presented in Tables \ref{tabla_02} and \ref{tabla_02A}.} The estimators were carried out using the software \cite{Rmanual}, and codes are shown in the supplementary file \ref{sf2}.
\begin{table}[H]
\begin{spacing}{1}
    \centering
    \begin{tabular}{|c|c|c|c|c|}
    \hline
       n& Par & $\rho=-0.2$ & $\rho=-0.4$ &$\rho=-0.8$\\ 
       \hline
       81 & \begin{tabular}{c}
       \\
           $\hat{\rho}$\\
           $\hat{\beta}_0$\\
           $\hat{\beta}_1$\\
           $\hat{\beta}_2$\\
           $\hat{\alpha}_0$\\
            $\hat{\alpha}_0$
      \end{tabular} &
      \begin{tabular}{ccc}
       Mean & Sd & Bias\\
       \hline
         -0.201 & 0.107 & 0.003 \\ 
5.367 & 1.093 & 1.68  \\ 
   -0.505 & 0.362 & 0.009  \\ 
  1.778 & 0.358 & 0.016 \\ 
0.320 & 0.263 & -0.359 \\ 
   0.345 & 0.120 & 0.151 \\ 
      \end{tabular}
      &
      \begin{tabular}{ccc}
       Mean & Sd & Bias\\
       \hline
         -0.403 & 0.103 & 0.008  \\ 
5.405 & 0.990 & 1.70 \\ 
   -0.485 & 0.331 & -0.031 \\ 
 1.767 & 0.359 & 0.009  \\ 
 0.315 & 0.248 & -0.369 \\ 
    0.352 & 0.111 & 0.173  \\ 
      \end{tabular}
      &
      \begin{tabular}{ccc}
       Mean & Sd & Bias\\
       \hline
          -0.789 & 0.058 & -0.014  \\ 
 5.389 & 0.791 & 1.69\\ 
 -0.516 & 0.350 & 0.031  \\ 
  1.744 & 0.380 & -0.003 \\ 
  0.315 & 0.263 & -0.369 \\ 
   0.351 & 0.118 & 0.172\\ 
      \end{tabular}\\
      \hline
       144 & \begin{tabular}{c}
       \\
           $\hat{\rho}$\\
           $\hat{\beta}_0$\\
           $\hat{\beta}_1$\\
           $\hat{\beta}_2$\\
           $\hat{\alpha}_0$\\
            $\hat{\alpha}_0$
      \end{tabular} &
      \begin{tabular}{ccc}
       Mean & Sd & Bias\\
       \hline
          -0.199 & 0.080 & -0.007 \\ 
  5.370 & 0.814 & 1.68 \\ 
 -0.503 & 0.244 & 0.006 \\ 
1.763 & 0.263 & 0.008  \\ 
  0.408 & 0.165 & -0.018  \\ 
 0.321 & 0.072 & 0.071\\ 
      \end{tabular}
      &
      \begin{tabular}{ccc}
       Mean & Sd & Bias\\
       \hline
         -0.395 & 0.079 & -0.012 \\ 
  5.356 & 0.691 & 1.67 \\ 
  -0.511 & 0.241 & 0.023 \\ 
  1.752 & 0.243 & 0.001 \\ 
  0.397 & 0.167 & -0.206  \\ 
  0.326 & 0.076 & 0.085  \\ 
      \end{tabular}
      &
      \begin{tabular}{ccc}
       Mean & Sd & Bias\\
       \hline
 -0.792 & 0.044 & -0.009 \\ 
  5.369 & 0.541 & 1.68  \\ 
-0.496 & 0.247 & -0.009\\ 
  1.733 & 0.252 & -0.010\\ 
 0.397 & 0.166 & -0.205 \\ 
 0.325 & 0.076 & 0.082 \\ 
      \end{tabular}\\
      \hline
       225 & \begin{tabular}{c}
       \\
           $\hat{\rho}$\\
           $\hat{\beta}_0$\\
           $\hat{\beta}_1$\\
           $\hat{\beta}_2$\\
           $\hat{\alpha}_0$\\
            $\hat{\alpha}_0$
      \end{tabular} &
      \begin{tabular}{ccc}
       Mean & Sd & Bias\\
       \hline
 -0.204 & 0.066 & 0.019 \\ 
 5.408 & 0.627 & 1.70\\ 
 -0.499 & 0.199 & -0.003 \\ 
 1.749 & 0.196 & -0.001 \\ 
 0.417 & 0.116 & -0.165\\ 
 0.323 & 0.052 & 0.076 \\ 
      \end{tabular}
      &
      \begin{tabular}{ccc}
       Mean & Sd & Bias\\
       \hline
 -0.399 & 0.059 & -0.004\\ 
 5.372 & 0.563 & 1.68 \\ 
  -0.511 & 0.189 & 0.022\\ 
 1.758 & 0.193 & 0.004\\ 
 0.441 & 0.122 & -0.119  \\ 
 0.312 & 0.054 & 0.004 \\  
      \end{tabular}
      &
      \begin{tabular}{ccc}
       Mean & Sd & Bias\\
       \hline
-0.798 & 0.033 & -0.003  \\ 
 5.410 & 0.433 & 1.70 \\ 
 -0.514 & 0.202 & 0.028  \\ 
 1.739 & 0.198 & -0.006 \\ 
0.432 & 0.125 & -0.137 \\ 
 0.317 & 0.056 & 0.055  \\ 
      \end{tabular}\\
      \hline
       400 & \begin{tabular}{c}
       \\
           $\hat{\rho}$\\
           $\hat{\beta}_0$\\
           $\hat{\beta}_1$\\
           $\hat{\beta}_2$\\
           $\hat{\alpha}_0$\\
            $\hat{\alpha}_0$
      \end{tabular} &
      \begin{tabular}{ccc}
       Mean & Sd & Bias\\
       \hline
-0.203 & 0.051 & 0.014 \\ 
 5.430 & 0.499 & 1.71 \\ 
 -0.493 & 0.142 & -0.014 \\ 
 1.748 & 0.145 & -0.001\\ 
0.460 & 0.087 & -0.079  \\ 
 0.310 & 0.038 & 0.034\\ 
      \end{tabular}
      &
      \begin{tabular}{ccc}
       Mean & Sd & Bias\\
       \hline
 -0.400 & 0.044 & 0.001  \\ 
 5.402 & 0.421 & 1.70\\ 
 -0.501 & 0.139 & 0.002 \\ 
  1.752 & 0.144 & 0.001 \\ 
0.456 & 0.086 & -0.088 \\ 
 0.311 & 0.038 & 0.037  \\  
      \end{tabular}
      &
      \begin{tabular}{ccc}
       Mean & Sd & Bias\\
       \hline
 0.023 & -0.405 & 0.001 \\ 
 5.381 & 0.311 & 1.69 \\ 
-0.503 & 0.143 & 0.005 \\ 
1.748 & 0.143 & -0.001 \\ 
 0.459 & 0.087 & -0.083\\ 
0.310 & 0.038 & 0.033  \\ 
      \end{tabular}\\
      
      \hline
    \end{tabular}
    \end{spacing}
    \caption{Results of the simulations of the parametric effects over mean and standard deviation models on a regular grid with $\beta_0=2$, $\beta_1$=-0.5, $\beta_2$=1.75, $\alpha_0$=0.5 and $\alpha_1$=0.3 for values of $\rho\in\{-0.2, -0.4, -0.8\}$}
    \label{tabla_02}
\end{table}
It is evident in Tables \ref{tabla_02} and \ref{tabla_02A} that the algorithm has a good capacity to capture the original parameters with which the data simulations were carried out. This performance is good for both the mean (including the parameter $\rho$) and the standard deviation. Likewise, it is highlighted that as the sample size increases, the bias in percentage terms is reduced, verifying the consistency of the estimators.

Regarding the parameter $\beta_0$, it can be seen that the estimate differs from the real value, because when a non-parametric component is estimated, within the reparametrization process developed, the means the variable is subtracted, to guarantee that the model has a single intercept and not one for each function, which means that $\beta_0$ cannot be estimated being it a common problem in GAMLSS models \citep{wood2017generalized}.

\begin{figure}[H]
\centering
\includegraphics[scale=0.8]{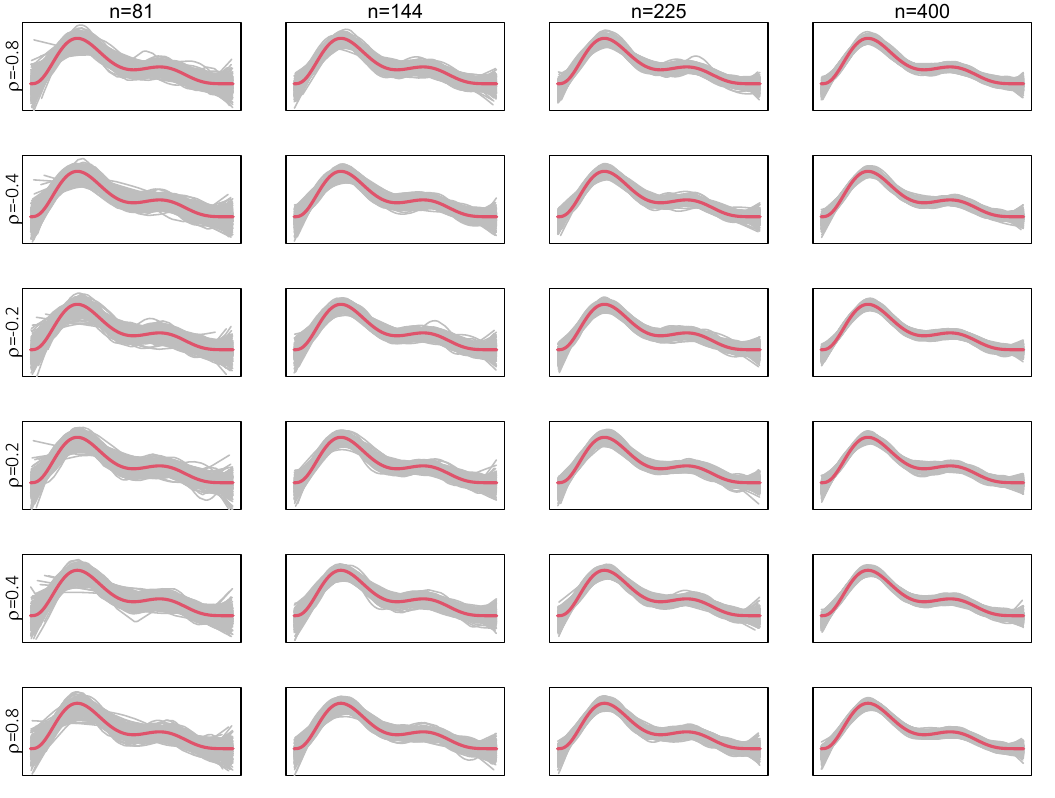}
\caption{Estimated functions for under different values of $\rho \in \{-0.8, -0.4, -0.2, 0.2, 0.4, 0.8\}$ and grid size $n=\{81,144, 225, 0.2, 0.4, 400 \}$ in the simulation. The mean estimated curve and curves from each simulation are presented.}
\label{fig_curvas1}
\end{figure}
On the other hand, analyzing Figure \ref{fig_curvas1}, it is observed that the estimated functions in the different simulations have a behavior similar to the real function. Also, as the sample size increases, the precision of the estimate improves. 
Furthermore, these results are maintained regardless of the value of $\rho$, which allows the conclusions about the parameters of interest to be generalizable without being affected by the value that the autoregressive spatial parameter may take.

\begin{figure}[H]
\centering
\includegraphics[scale=0.52]{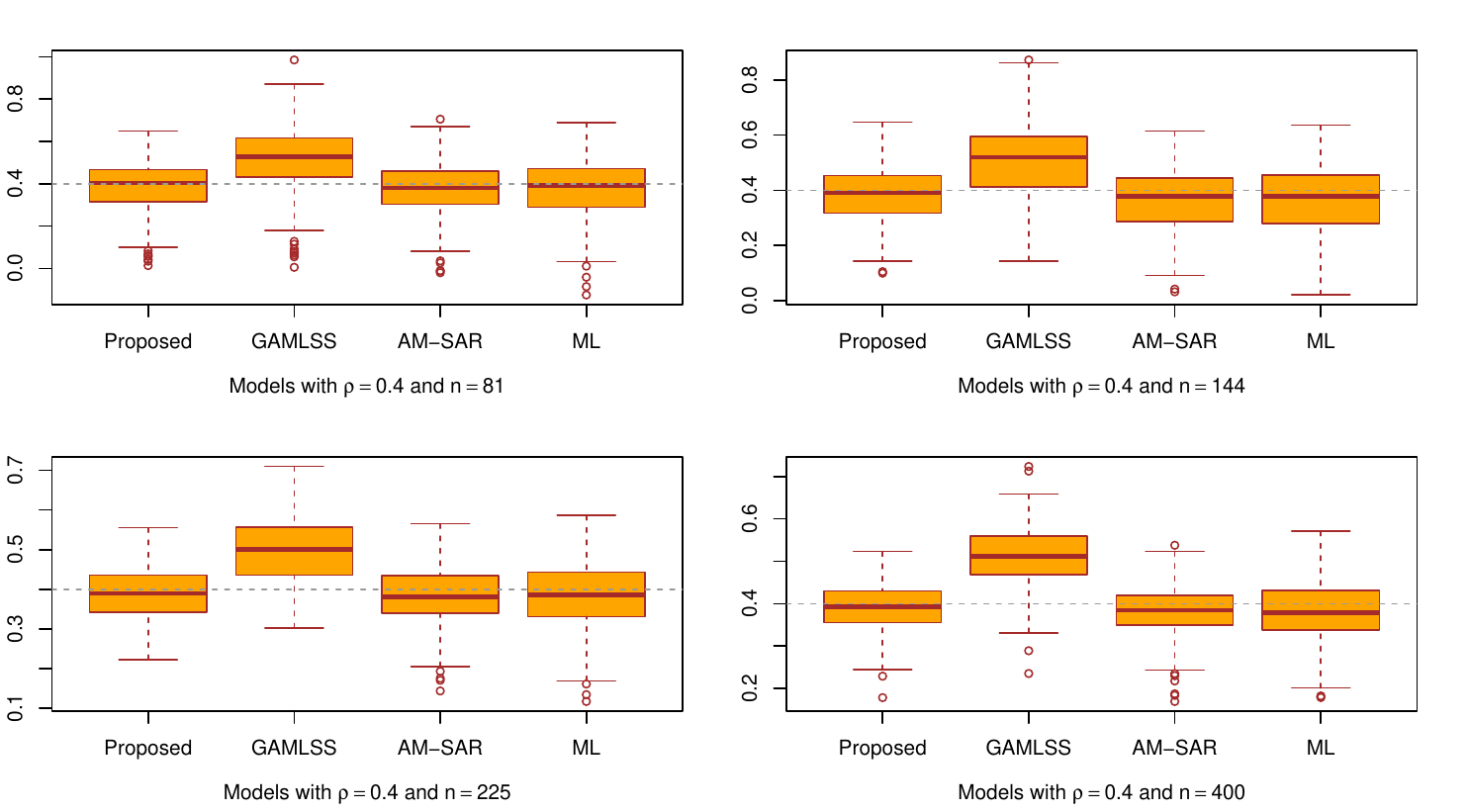}
\caption{$\hat{\rho}$ for the different models adjusted for the simulation in a regular grid with $\rho=0.4$.}
\label{02_81_rho}
\end{figure}

\begin{table}[H]
    \centering
    \begin{tabular}{|c|c|c|c|c|}
    \hline
       n& Par & $\rho=0.2$ & $\rho=0.4$ &$\rho=0.8$\\ 
       \hline
       81 & \begin{tabular}{c}
       \\
           $\hat{\rho}$\\
           $\hat{\beta}_0$\\
           $\hat{\beta}_1$\\
           $\hat{\beta}_2$\\
           $\hat{\alpha}_0$\\
            $\hat{\alpha}_0$
      \end{tabular} &
      \begin{tabular}{ccc}
       Mean & Sd & Bias\\
       \hline
0.178 & 0.110 & -0.109 \\ 
5.669 & 1.370 & 1.83  \\ 
-0.490 & 0.340 & -0.019 \\ 
1.728 & 0.335 & -0.013 \\ 
0.332 & 0.237 & -0.335  \\ 
0.341 & 0.107 & 0.137  \\ 
      \end{tabular}
      &
      \begin{tabular}{ccc}
       Mean & Sd & Bias\\
       \hline
 0.369 & 0.108 & -0.077\\ 
 5.859 & 1.779 & 1.92  \\ 
 -0.519 & 0.362 & 0.038  \\ 
1.754 & 0.363 & 0.002\\ 
 0.333 & 0.268 & -0.335 \\ 
 0.341 & 0.117 & 0.136  \\ 
      \end{tabular}
      &
      \begin{tabular}{ccc}
       Mean & Sd & Bias\\
       \hline
0.778 & 0.062 & -0.027 \\ 
 6.337 & 2.772 & 2.16 \\ 
-0.514 & 0.315 & 0.028 \\ 
 1.739 & 0.380 & -0.006 \\ 
 0.317 & 0.275 & -0.366 \\ 
0.350 & 0.118 & 0.165 \\ 
      \end{tabular}\\
      \hline
       144 & \begin{tabular}{c}
       \\
           $\hat{\rho}$\\
           $\hat{\beta}_0$\\
           $\hat{\beta}_1$\\
           $\hat{\beta}_2$\\
           $\hat{\alpha}_0$\\
            $\hat{\alpha}_0$
      \end{tabular} &
      \begin{tabular}{ccc}
       Mean & Sd & Bias\\
       \hline
 0.191 & 0.081 & -0.046  \\ 
 5.494 & 1.020 & 1.74  \\ 
  -0.513 & 0.250 & 0.026\\ 
 1.746 & 0.266 & -0.002  \\ 
 0.395 & 0.162 & -0.209  \\ 
0.327 & 0.073 & 0.088  \\ 
      \end{tabular}
      &
      \begin{tabular}{ccc}
       Mean & Sd & Bias\\
       \hline
 0.382 & 0.075 & -0.043\\ 
 5.644 & 1.206 & 1.82 \\ 
-0.506 & 0.244 & 0.013  \\ 
 1.753 & 0.252 & 0.001 \\ 
 0.397 & 0.162 & -0.206  \\ 
 0.327 & 0.075 & 0.090  \\ 
      \end{tabular}
      &
      \begin{tabular}{ccc}
       Mean & Sd & Bias\\
       \hline
0.785 & 0.046 & -0.012 \\ 
6.069 & 2.083 & 2.035 \\ 
 -0.511 & 0.240 & 0.021 \\ 
 1.760 & 0.241 & 0.005 \\ 
 0.397 & 0.162 & -0.206 \\ 
 0.328 & 0.073 & 0.095 \\
      \end{tabular}\\
      \hline
       225 & \begin{tabular}{c}
       \\
           $\hat{\rho}$\\
           $\hat{\beta}_0$\\
           $\hat{\beta}_1$\\
           $\hat{\beta}_2$\\
           $\hat{\alpha}_0$\\
            $\hat{\alpha}_0$
      \end{tabular} &
      \begin{tabular}{ccc}
       Mean & Sd & Bias\\
       \hline
 0.192 & 0.068 & -0.041 \\ 
5.458 & 0.816 & 1.72\\ 
-0.503 & 0.204 & 0.005\\ 
1.757 & 0.194 & 0.004  \\ 
0.434 & 0.117 & -0.131  \\ 
0.315 & 0.053 & 0.049  \\
      \end{tabular}
      &
      \begin{tabular}{ccc}
       Mean & Sd & Bias\\
       \hline
0.391 & 0.059 & -0.021 \\ 
 5.540 & 0.947 & 1.77 \\ 
-0.493 & 0.184 & -0.013 \\ 
1.741 & 0.188 & -0.005 \\ 
0.434 & 0.124 & -0.132 \\ 
 0.316 & 0.054 & 0.053\\
      \end{tabular}
      &
      \begin{tabular}{ccc}
       Mean & Sd & Bias\\
       \hline
0.790 & 0.033 & -0.013 \\ 
 5.854 & 1.490 & 1.92 \\ 
 -0.504 & 0.190 & 0.007 \\ 
 1.754 & 0.208 & 0.002 \\ 
 0.438 & 0.117 & -0.124 \\ 
 0.316 & 0.052 & 0.053 \\
      \end{tabular}\\
      \hline
       400 & \begin{tabular}{c}
       \\
           $\hat{\rho}$\\
           $\hat{\beta}_0$\\
           $\hat{\beta}_1$\\
           $\hat{\beta}_2$\\
           $\hat{\alpha}_0$\\
            $\hat{\alpha}_0$
      \end{tabular} &
      \begin{tabular}{ccc}
       Mean & Sd & Bias\\
       \hline
0.195 & 0.048 & -0.023  \\ 
 5.471 & 0.598 & 1.73 \\ 
 -0.503 & 0.148 & 0.006 \\ 
 1.752 & 0.134 & 0.001 \\ 
 0.463 & 0.087 & -0.074  \\ 
0.308 & 0.039 & 0.026  \\ 
      \end{tabular}
      &
      \begin{tabular}{ccc}
       Mean & Sd & Bias\\
       \hline
 0.395 & 0.047 & -0.013  \\ 
 5.479 & 0.737 & 1.73 \\ 
 -0.504 & 0.142 & 0.008 \\ 
 1.745 & 0.143 & -0.002 \\ 
  0.465 & 0.083 & -0.071 \\ 
 0.309 & 0.037 & 0.028 \\ 
      \end{tabular}
      &
      \begin{tabular}{ccc}
       Mean & Sd & Bias\\
       \hline
 0.793 & 0.026 & -0.008 \\ 
  5.705 & 1.155 & 1.85 \\ 
  -0.504 & 0.147 & 0.008 \\ 
 1.746 & 0.138 & -0.002 \\ 
0.466 & 0.091 & -0.068 \\ 
 0.307 & 0.040 & 0.024 \\ 
      \end{tabular}\\
      
      \hline
    \end{tabular}
    \caption{Results of the simulations of the parametric effects over mean and standard deviation models on a regular grid with $\beta_0=2$, $\beta_1$=-0.5, $\beta_2$=1.75, $\alpha_0$=0.5 and $\alpha_1$=0.3 for values of $\rho\in\{0.2, 0.4, 0.8\}$.}
    \label{tabla_02A}
\end{table}
\begin{figure}[H]
\centering
\includegraphics[scale=0.52]{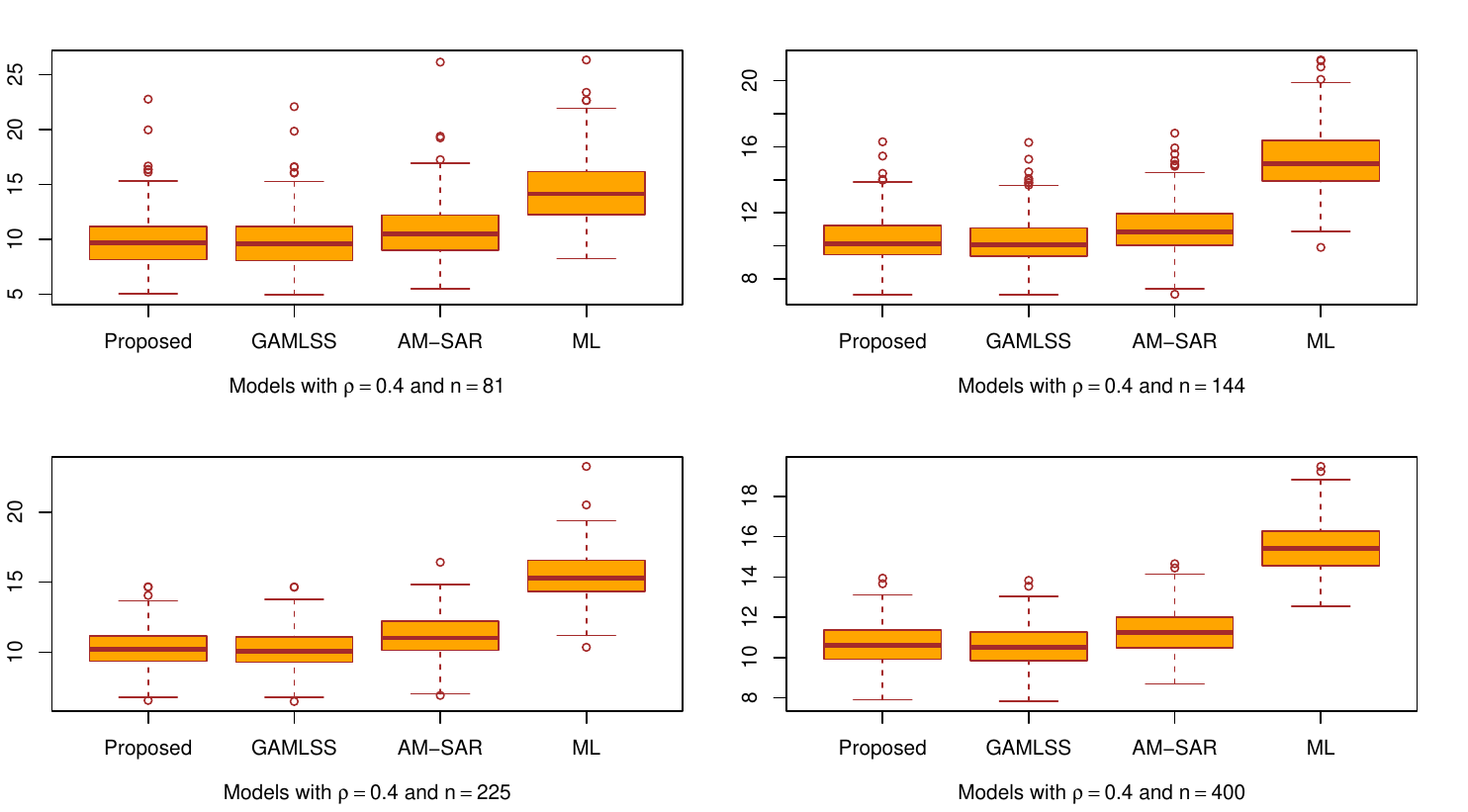}
\caption{MSE for the different models adjusted for the simulation in a regular grid with $\rho=0.4$.}
\label{02_81_mse}
\end{figure}
\textcolor{red}{Additionally, Figure \ref{02_81_mse} shows the results in terms of the mean square error (MSE), where those that incorporate the estimation of the non-parametric component perform considerably better than the classical linear proposal. Furthermore, Figure \ref{02_81_rho} shows that the estimate of the parameter $\rho$ is better in terms of bias and variance in the proposed methodology with respect to the other methodologies.\\
Finally, from Tables \ref{tabla_02} and \ref{tabla_02A} and the results of supplementary file \ref{sf1} (results of the estimators and MSE), it is observed that as the size increases or sample ($n=144$, $n=225$ and $n=400$), the precision of the estimates improves considerably, which is associated with the consistency of the estimators; with the exception of the GAMLSS with spatial lag where the $\rho$ does not tend to its true value despite the increase in the number of observations. Likewise, the results obtained with $n=81$ for H-AM-SAR are maintained with the different values of $n$, showing lower levels of variability in the estimates of the parameters associated with the mean, with respect to other methods such as ML or AM-SAR, and estimates of $\alpha_0$ and $\alpha_1$ that are adequate. This behavior guarantees that the estimate obtained by the methodology proposed in this paper has more desirable statistical properties than with the usual methods, which will lead to better inference from the observed data.}

\begin{figure}[H]
\centering
\includegraphics[scale=0.58]{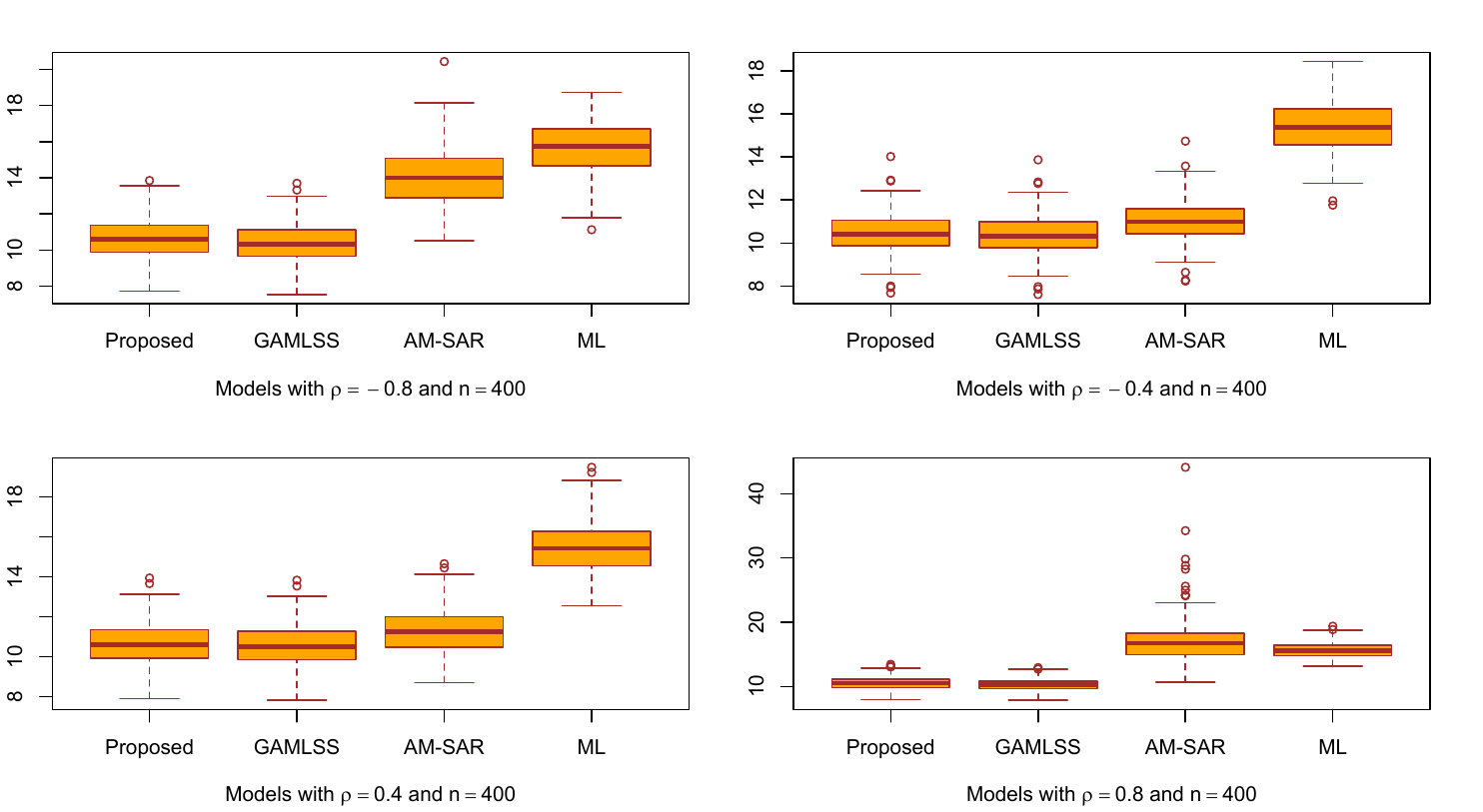}
\caption{MSE for the different models adjusted for the simulation in a spatial points grid with $n=400$.}
\label{Cundi_mse}
\end{figure}

\textcolor{red}{
\subsection{Analysis in spatial points}
How later an application is made on a set of spatial points in Bogotá, here a simulation exercise similar to the previous one but without considering a grid, but rather a set of spatial points is carried out with the same parameters as the previous simulation and $n=81, 144, 225, 400$ points.
Furthermore, to incorporate the spatial effects, an inverse squared distance matrix was used, which was standardized to be able to incorporate in each spatial unit the effect of proximity to the other homes $(w_{ij}=1/d_{ij} ^2)$. 
\begin{figure}[H]
\centering
\includegraphics[scale=0.35]{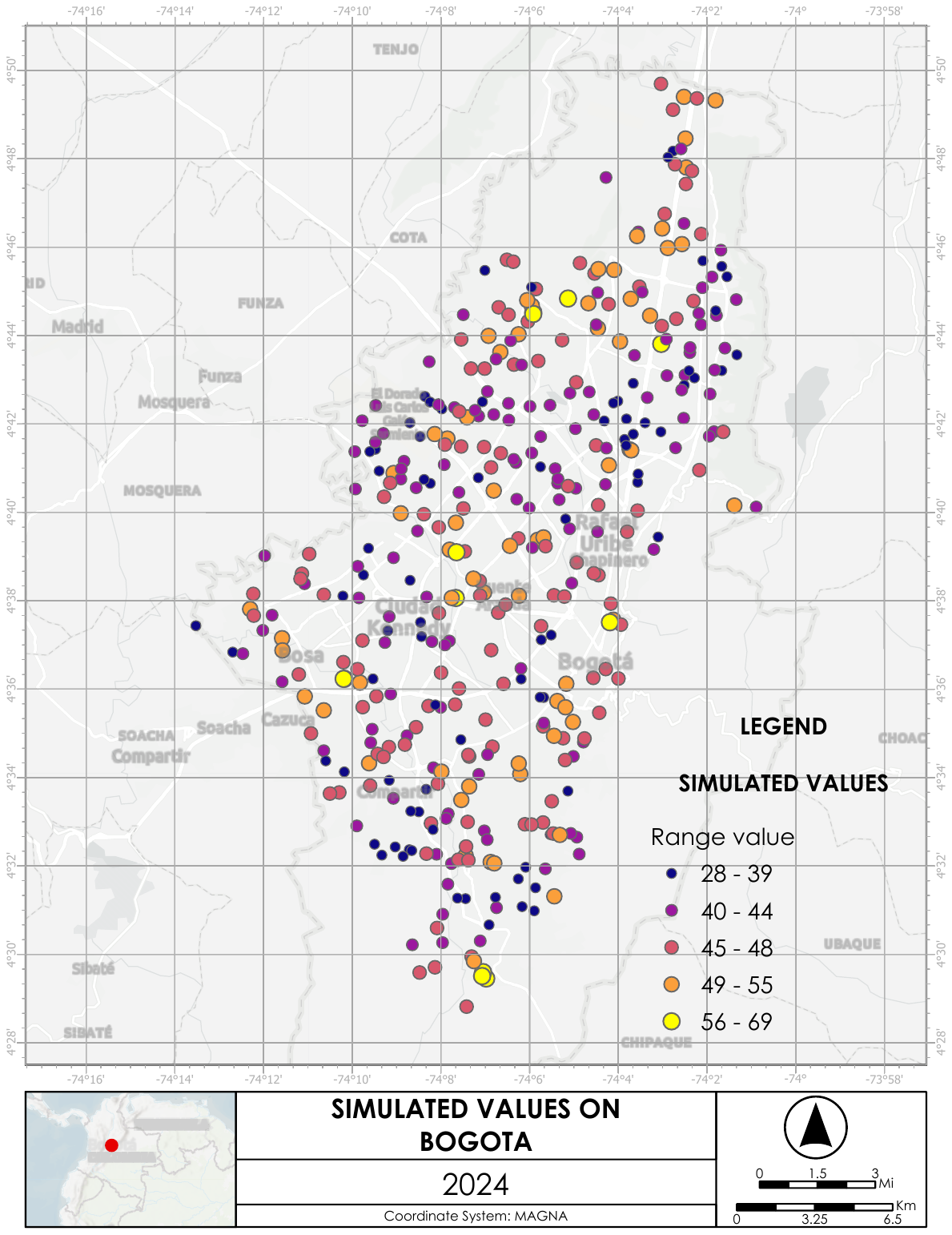}
\caption{Simulated values on spatial points of Bogotá with $n=400$ and $\rho=0.2$.}
\label{mapa_puntos}
\end{figure}
Figure \ref{mapa_puntos} shows one of the simulation results with $n=400$ and a spatial correlation coefficient $\rho=0.2$. Figure \ref{Cundi_mse} shows the results in terms of the mean square error (MSE), where those that incorporate the estimation of the non-parametric component perform considerably better than the classical linear proposal. Furthermore, the MSE behaves similarly to the results obtained in the regular grid, and in supplementary file 1, it is observed that this behavior is maintained with different values of $\rho$.}
\textcolor{red}{It is evident in Table \ref{tabla_SP_02} that the algorithm has a good capacity to capture the original parameters with which the data simulations were carried out, and with a behavior similar to the regular grid. This performance is good for both the mean (including the parameter $\rho$) and the standard deviation. Likewise, it is highlighted that as the sample size increases, the bias is reduced verifying the consistency of the estimators. Looking at tables in Supplementary File 1, similar results were obtained using our proposed model than the other models.}
\begin{table}[H]
\begin{spacing}{1}
    \centering
    \begin{tabular}{|c|c|c|c|c|}
    \hline
       n& Par & $\rho=0.2$ & $\rho=0.4$ &$\rho=0.8$\\ 
       \hline
       81 & \begin{tabular}{c}
       \\
           $\hat{\rho}$\\
           $\hat{\beta}_0$\\
           $\hat{\beta}_1$\\
           $\hat{\beta}_2$\\
           $\hat{\alpha}_0$\\
            $\hat{\alpha}_0$
      \end{tabular} &
      \begin{tabular}{ccc}
       Mean & Sd & Bias\\
       \hline
  0.168 & 0.139 & -0.032 \\ 
  5.799 & 1.726 & 3.799 \\ 
  -0.525 & 0.342 & -0.025 \\ 
  1.736 & 0.373 & -0.014 \\ 
  0.308 & 0.251 & -0.192 \\ 
  0.349 & 0.110 & 0.049 \\ 
      \end{tabular}
      &
      \begin{tabular}{ccc}
       Mean & Sd & Bias\\
       \hline
 0.386 & 0.126 & -0.014 \\ 
  5.575 & 1.995 & 3.575 \\ 
  -0.472 & 0.322 & 0.028 \\ 
  1.758 & 0.374 & 0.008 \\ 
  0.343 & 0.253 & -0.157 \\ 
  0.338 & 0.113 & 0.038 \\ 
      \end{tabular}
      &
      \begin{tabular}{ccc}
       Mean & Sd & Bias\\
       \hline
0.767 & 0.073 & -0.033 \\ 
  6.909 & 3.313 & 4.909 \\ 
  -0.443 & 0.332 & 0.057 \\ 
  1.734 & 0.357 & -0.016 \\ 
  0.341 & 0.236 & -0.159 \\ 
  0.340 & 0.109 & 0.040 \\ 
      \end{tabular}\\
      \hline
       144 & \begin{tabular}{c}
       \\
           $\hat{\rho}$\\
           $\hat{\beta}_0$\\
           $\hat{\beta}_1$\\
           $\hat{\beta}_2$\\
           $\hat{\alpha}_0$\\
            $\hat{\alpha}_0$
      \end{tabular} &
      \begin{tabular}{ccc}
       Mean & Sd & Bias\\
       \hline
 0.176 & 0.108 & -0.024 \\ 
  5.697 & 1.261 & 3.697 \\ 
  -0.473 & 0.236 & 0.027 \\ 
  1.713 & 0.240 & -0.037 \\ 
  0.382 & 0.170 & -0.118 \\ 
  0.333 & 0.073 & 0.033 \\ 
      \end{tabular}
      &
      \begin{tabular}{ccc}
       Mean & Sd & Bias\\
       \hline
0.381 & 0.105 & -0.019 \\ 
  5.715 & 1.662 & 3.715 \\ 
  -0.499 & 0.232 & 0.001 \\ 
  1.718 & 0.267 & -0.032 \\ 
  0.404 & 0.166 & -0.096 \\ 
  0.324 & 0.073 & 0.024 \\ 
      \end{tabular}
      &
      \begin{tabular}{ccc}
       Mean & Sd & Bias\\
       \hline
0.779 & 0.060 & -0.021 \\ 
  6.223 & 2.699 & 4.223 \\ 
  -0.497 & 0.254 & 0.003 \\ 
  1.807 & 0.259 & 0.057 \\ 
  0.404 & 0.180 & -0.096 \\ 
  0.324 & 0.080 & 0.024 \\ 
      \end{tabular}\\
      \hline
       225 & \begin{tabular}{c}
       \\
           $\hat{\rho}$\\
           $\hat{\beta}_0$\\
           $\hat{\beta}_1$\\
           $\hat{\beta}_2$\\
           $\hat{\alpha}_0$\\
            $\hat{\alpha}_0$
      \end{tabular} &
      \begin{tabular}{ccc}
       Mean & Sd & Bias\\
       \hline
 0.202 & 0.066 & 0.002 \\ 
  5.406 & 0.817 & 3.406 \\ 
  -0.500 & 0.194 & -0.000 \\ 
  1.754 & 0.188 & 0.004 \\ 
  0.428 & 0.128 & -0.072 \\ 
  0.319 & 0.059 & 0.019 \\ 
      \end{tabular}
      &
      \begin{tabular}{ccc}
       Mean & Sd & Bias\\
       \hline
0.388 & 0.067 & -0.012 \\ 
  5.560 & 1.062 & 3.560 \\ 
  -0.510 & 0.192 & -0.010 \\ 
  1.762 & 0.191 & 0.012 \\ 
  0.433 & 0.115 & -0.067 \\ 
  0.314 & 0.055 & 0.014 \\ 
      \end{tabular}
      &
      \begin{tabular}{ccc}
       Mean & Sd & Bias\\
       \hline
0.788 & 0.038 & -0.012 \\ 
  5.918 & 1.666 & 3.918 \\ 
  -0.511 & 0.178 & -0.011 \\ 
  1.760 & 0.197 & 0.010 \\ 
  0.431 & 0.118 & -0.069 \\ 
  0.320 & 0.053 & 0.020 \\ 
      \end{tabular}\\
      \hline
       400 & \begin{tabular}{c}
       \\
           $\hat{\rho}$\\
           $\hat{\beta}_0$\\
           $\hat{\beta}_1$\\
           $\hat{\beta}_2$\\
           $\hat{\alpha}_0$\\
            $\hat{\alpha}_0$
      \end{tabular} &
      \begin{tabular}{ccc}
       Mean & Sd & Bias\\
       \hline
0.188 & 0.075 & -0.012 \\ 
  5.514 & 0.901 & 3.514 \\ 
  -0.493 & 0.156 & 0.007 \\ 
  1.750 & 0.156 & -0.000 \\ 
  0.458 & 0.078 & -0.042 \\ 
  0.311 & 0.034 & 0.011 \\ 
      \end{tabular}
      &
      \begin{tabular}{ccc}
       Mean & Sd & Bias\\
       \hline
 0.390 & 0.060 & -0.010 \\ 
  5.536 & 0.948 & 3.536 \\ 
  -0.504 & 0.145 & -0.004 \\ 
  1.744 & 0.137 & -0.006 \\ 
  0.462 & 0.085 & -0.038 \\ 
  0.310 & 0.039 & 0.010 \\ 
      \end{tabular}
      &
      \begin{tabular}{ccc}
       Mean & Sd & Bias\\
       \hline
 0.785 & 0.044 & -0.015 \\ 
  6.055 & 1.987 & 4.055 \\ 
  -0.503 & 0.143 & -0.003 \\ 
  1.756 & 0.145 & 0.006 \\ 
  0.468 & 0.087 & -0.032 \\ 
  0.307 & 0.038 & 0.007 \\ 
      \end{tabular}\\
      
      \hline
    \end{tabular}
    \end{spacing}
    \caption{Results of the simulations of the parametric effects over mean and standard deviation models on spatial points with $\beta_0=2$, $\beta_1$=-0.5, $\beta_2$=1.75, $\alpha_0$=0.5 and $\alpha_1$=0.3 for values of $\rho\in\{0.2, 0.4, 0.8\}$}
    \label{tabla_SP_02}
\end{table}
\begin{figure}[H]
\centering
\includegraphics[scale=0.58]{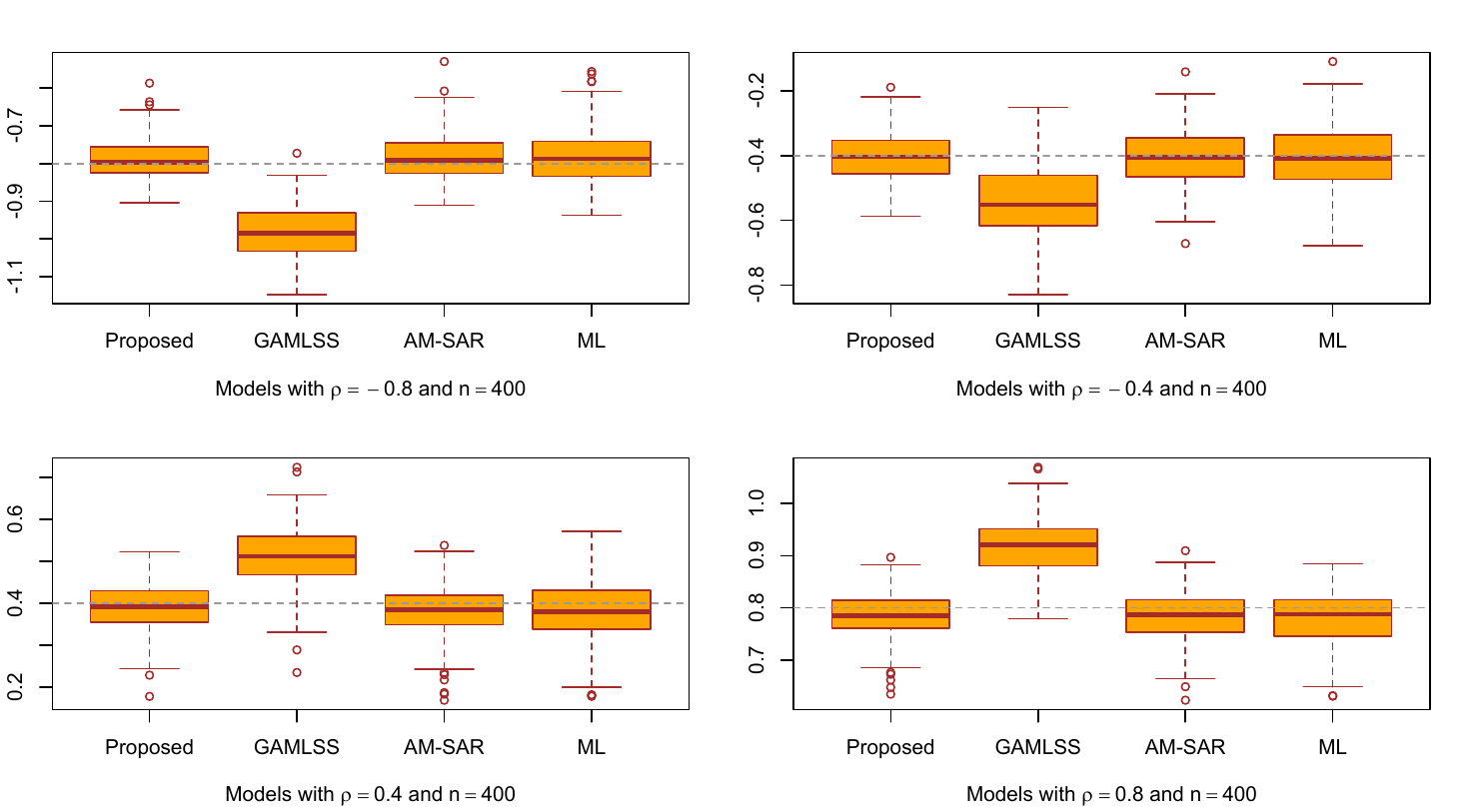}
\caption{$\hat{\rho}$ for the different models adjusted in the simulation using a spatial points grid with $n=400$. The dashed gray line indicates the true value of parameter $\rho$.}
\label{SP_rho}
\end{figure}
\textcolor{red}{Furthermore, the estimate values of $\rho$ are shown in Figure \ref{SP_rho} demonstrating the advantage over the GAMLSS model, similar to the AM-SAR and less dispersed than the ML, which combined with the MSE proves that the proposed model is more robust to estimate spatial correlations.}
\section{Application}
\begin{figure}[H]
    \centering
    \includegraphics[scale=0.35]{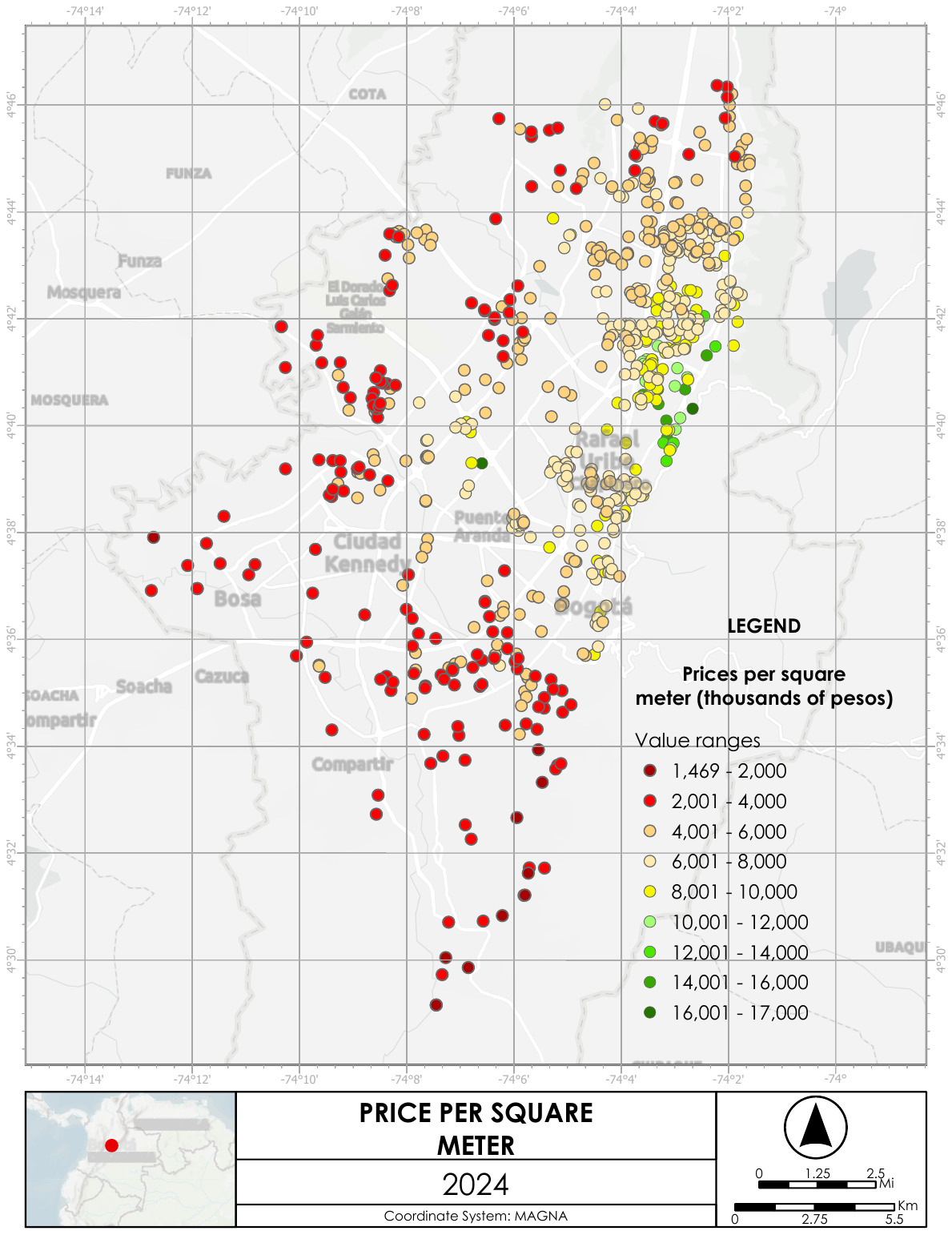}
    \caption{Prices per square meter of the new housing projects in Bogotá city (Colombia).}
    \label{mapa_bogota}
\end{figure}

From the building census carried out by Camacol in Bogotá and Cundinamarca within the framework of its georeferenced information system, information was obtained on 715 new housing projects that as of November 2019 had homes available for sale (see Figure \ref{mapa_bogota}). For these projects, the distance to Transmilenio stations (a transportation system), public parks, and main roads were calculated to determine if these variables have a non-linear relationship that positively or negatively impacts housing prices. Additionally, the database has information on the delivery condition, the socioeconomic stratum, and the number of bathrooms and bedrooms in each of the homes.
\textcolor{red}{The area variable is included in a non-parametric effect, considering non-linear relationships with the price per square meter \citep{feng2021non}. Likewise, distances to roads, Transmilenio stations, and parks can show non-linear behavior in relation to housing price, as \citet{shimizu2014nonlinearity} demonstrated. The other regressors are included as linear predictors, taking into account that they are qualitative attributes and discrete values such as bathrooms and bedrooms; therefore, the following hedonic pricing models are proposed:}
\begin{align*}
    {\ln(P_i)}&={\rho}\sum_{j=1}^n w_{ij}\ln(P_i)+{\beta}_0+{\beta}_1(ST 3_i)+{\beta}_2(ST 4_i)+\\
    &{\beta}_3(ST 5_i)+{\beta}_4(ST6_i)+{\beta}_5(GW_i) +{\beta}_6(UH_i)+{\beta}_7(BD_i)+\\
    &{\beta}_8(BT_i)+
    {f}_1(Area_i)+{f}_2(TM_i)+{f}_3(PK_i)+{f}_4(RD_i)\\
    {\ln(\sigma_i)}&={\alpha}_0+{\alpha}_1(ST 3_i)+{\alpha}_2(ST 4_i)+{\alpha}_3(ST5_i)+{\alpha}_4(ST 6_i)+\\
&{\alpha}_5(BT_i)+{g}_1(Area)+{g}_2(TM)+{g}_3(PK)+{g}_4(RD)
\end{align*}

The definitions of the variables are: i) P: price per square meter, ii) TM: distance to the nearest Transmilenio station (meters), iii) ST: socioeconomic stratum of the project (factor), iv) Bedrooms: number of bedrooms in the house, v) Area: area at house (square meters), vi) PK: distance to nearest park (meters), vii) RD: distance to main roads (meters), viii) BT: number of bathrooms in the house, ix) UH: corresponds to the finishes in the house, x) GW: an indicator of unfinished in the house.
As is common in this kind of model, a natural logarithm was applied to the price per square meter variable, so that the estimated effects are semi elasticities and make sense in economic terms. This is common when the distribution of prices is assumed to follow a log-normal.
Furthermore, to incorporate the spatial effects, an inverse squared distance matrix was used, which was standardized to be able to incorporate in each spatial unit, the effect of proximity to the other homes $(w_{ij}=1/d_{ij} ^2)$.
Before starting to estimate the models, an exploratory analysis of spatial data (AEDE) is carried out to establish whether the observations present some type of spatial dependence.
\begin{figure}[H]
    \centering
    \includegraphics[scale=0.53]{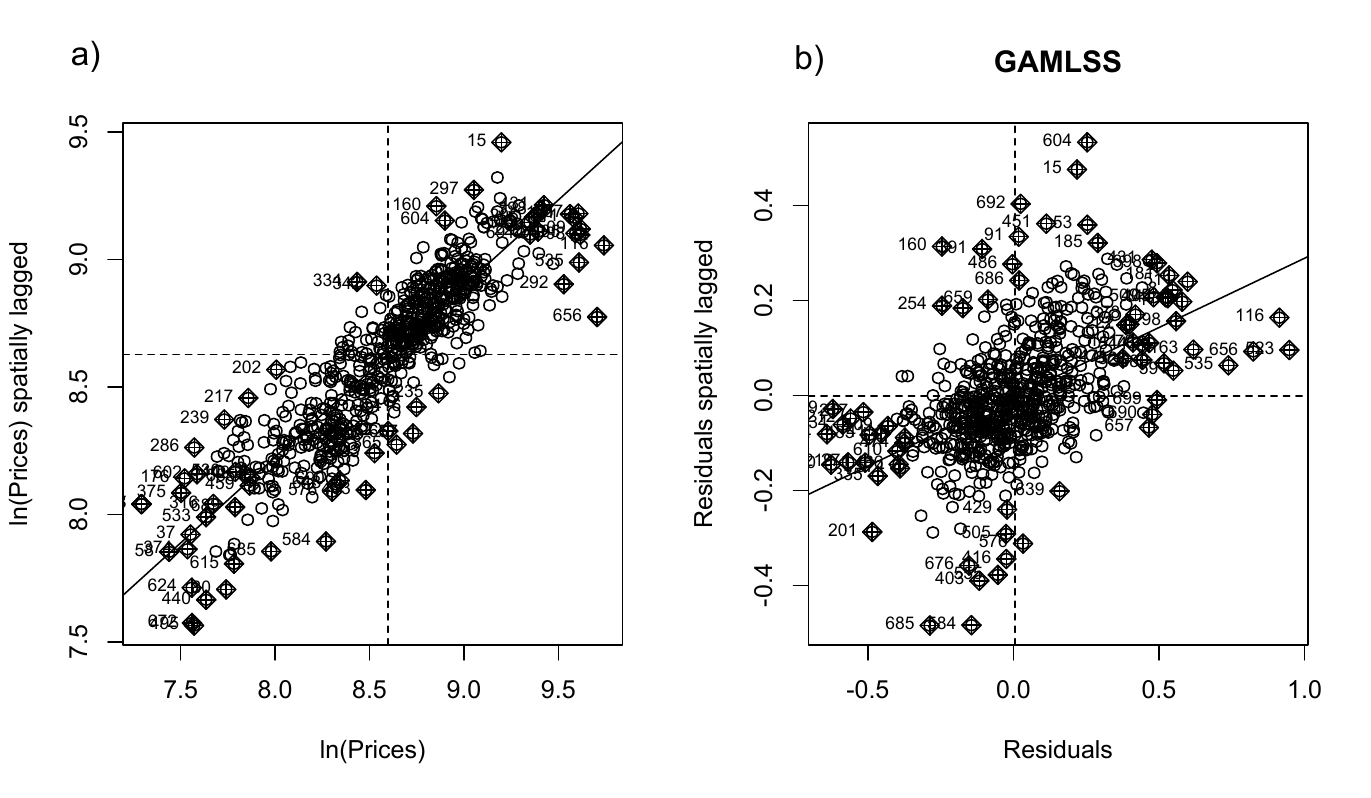}
    \caption{Moran's I plot on the residuals obtained from the model built for the hedonic prices of Bogotá city only with GAMLSS: a) with respect to the ln(Price) and, b) with respect to the same residuals}
    \label{moran_inicial}
\end{figure}

Figure \ref{moran_inicial} shows Moran's I plot for the logarithm of prices (dependent variable) and the residuals obtained after estimating a GAMLSS model for mean and variance that did not include the spatial lag. A clear spatial dependence is evident, which was confirmed by calculating Moran's I, obtaining significant results for the two variables (0.6729 for ln(price) and 0.2916 for the residuals).
Consequently, the existing spatial dependence between observations is confirmed, and it is inferred that traditional methods such as OLS or GAMLSS, which do not include modeling of the spatial autoregressive component, will not be suitable for this situation. 
{
\begin{table}[H]
\centering
\footnotesize
\begin{tabular}{|ccccc|}
  \hline
Parameter& Proposed & GAMLSS & AM-SAR & ML \\ 
  \hline
 \multicolumn{5}{|c|}{Mean($\mu$)}\\
 \hline
Intercept & 2.9823  & 3.1021 & 3.1994 & 2.9230 \\ 
  Area & &  &  & 0.0008 \\ 
  BT &0.01747  & 0.0175  & 0.0230 & 0.0187  \\ 
  BD &-0.0527  & -0.0534 & -0.0566 & -0.0729   \\ 
  TM &  &  &  & -0.00003\\ 
  RD & &  &  & 0.00002 \\ 
  PK &  &  &  & 0.00006  \\ 
  GW & -0.1538 & -0.1553  & -0.1427 & -0.1081  \\ 
  UH & -0.3753& -0.3768  &-0.2797  &-0.2555  \\ 
  ST3 & 0.1929   & 0.1946 & 0.1907 & 0.1916 \\ 
  ST4 &  0.3315 &  0.3367 & 0.3242 & 0.3191 \\ 
  ST5 & 0.3988 & 0.4056  & 0.4046  & 0.3837 \\ 
  ST3 & 0.4926  & 0.5023& 0.4790  & 0.4732 \\ 
 \hline
 \multicolumn{5}{|c|}{Desviaci\'{o}n est\'{a}ndar  ($\sigma$)}\\
 \hline
 Intercept &-1.37670 &-1.38411 & & \\
  BT & -0.2248 & -0.2230  &  & \\ 
  ST3 & 0.0232 & 0.0303&  & \\ 
  ST4 & -0.3209 &   -0.3124 & & \\ 
  ST5 & -0.3237 & -0.3125  &   & \\ 
  ST3 & -0.1569& -0.1464&   &  \\ 
  \hline
  $\rho$ & 0.6282 & 0.6140 & 0.6187 & 0.6363  \\
  \hline
  MSE & 0.0230 & 0.0272 & 0.0436 & 0.0288  \\ 
   \hline
   \end{tabular}
   \caption{Results of the estimation of the parametric effects over mean and standard
deviation for prices of four models}
   \label{resulBogota}
\end{table}
}
\textcolor{red}{The same specification for the mean was carried out with the ML, AM-SAR, and GAMLSS model using $Wy$ as an explanatory variable, and the results are shown in Table \ref{resulBogota}. This table shows the best performance of our proposed model in terms of MSE. Furthermore, according to the simulation results, the $\rho$ estimator with the least bias is provided by our model. Therefore, the proposed methodology becomes relevant.} 

\textcolor{red}{To analyze the effect of covariates on housing prices, the value of the estimator is incorrect because the parameters of a SAR model cannot be interpreted as marginal effects. Therefore, the direct, indirect (spillover), and total average effects must be calculated.
The direct effect is represented by the mean of the diagonal terms of the partial derivative matrix, $\pmb{S}_k$, defined as:
\begin{equation}\label{sk_dit}
    \pmb{S}_k = \beta_k(\mathbf{I}-\rho\pmb{W})^{-1}
\end{equation}
The indirect effect is defined as the mean of the off-diagonal elements in each row (or column) of the same matrix, $\pmb{S}_k$. The total effect is represented as the sum of the direct and indirect effects \citep{lesage2009introduction}.}

In Table \ref{resulBog}, the estimates for the parametric terms of the models are presented.
Thus, by observing Table \ref{resulBog}, the following is observed:
the estimates of the linear parameters obtained from the proposed methodology have results consistent with those obtained by \cite{delgado2021determinants}, who did not include the variance component within the estimation. \textcolor{red}{ For practicality, only the total impact on the dependent variable will be analyzed. By multiplying the total impact by the mean of the corresponding explanatory variable, the magnitude could be interpreted as an elasticity, as done by \citet{kim2003measuring} to determine the benefits of air quality on housing prices.}

For each additional bathroom, the price per square meter will increase by 4.73\%. Furthermore, for each additional room, the price per square meter is expected to reduce by 14.25\%, keeping all other variables constant. Given this result, it is important to emphasize the demographic change that Bogotá city is experiencing, where homes are increasingly smaller and require fewer rooms. Therefore, the offer is mainly concentrated on this type of property with few rooms and only twelve projects that offered homes with more than three bedrooms within the database.
\textcolor{red}{
\begin{table}[H]
     \centering
 \begin{tabular}{c|c|c|c|c|}
Variable & Estimate & Direct effect & Indirect effect& Total effect \\
\hline
        Bathrooms  & 0.0175 & 0.0195 & 0.0278& 0.0473 \\
        Bedrooms &  -0.0526 & -0.0588 & -0.0837& -0.1425 \\
        Gray Work & -0.1536 & -0.1716 & -0.2442 & -0.4158 \\
        Unfinished Home & -0.3751 & -0.4190  & -0.5964& -1.0154  \\
        Stratum 3 &  0.1927 & 0.2153 & 0.3064& 0.5217 \\
        Stratum 4 & 0.3306 & 0.3693 & 0.5256 & 0.8949 \\
        Stratum 5 &  0.3977 & 0.4443& 0.6323 & 1.0766 \\
        Stratum 6 & 0.4910 & 0.5486& 0.7808 & 1.3294 \\
        \hline
        \end{tabular}
     \caption{Estimation results for linear parameters in the hedonic price model Bogotá city with semiparametric SAR with non-homogeneous variance}
     \label{resulBog}
\end{table}
}
The fact that a home is delivered in gray construction, reduces on average its price per square meter by 41.58\%, compared to a property that is delivered finished. Likewise, when homes are delivered in black construction, they have an average price per square meter 101.5\% lower than that of a finished property, keeping all other variables constant.

In Bogotá city, the socioeconomic strata are well-defined according to the location of the properties, where there is a relationship between this variable and the housing prices. This relationship is reflected in the model, according to which, as the stratum increases, the price per square meter increases: a house in stratum three has on average a price per square meter 52\% higher than a house in stratum two, keeping all other variables constant. Analogously, a house in stratum four has on average a price per square meter 89\% higher than a house in stratum two, keeping all other variables constant, and so on for each stratum compared to stratum two.

The estimate of $\rho$ is above 0.6, which means that there is considerable spatial dependence, where high-priced-per-square-meter homes are surrounded by other high-priced homes, as evident in the map (see Figure \ref {mapa_bogota}). Additionally, the number of bathrooms has an inverse relationship concerning the variability of the price per square meter of the properties. An increase in price variability per square meter is evident in stratum 3 homes concerning stratum 2. The above is largely explained by the fact that this stratum offers so much social interest housing (SIH) as not SIH, which causes price dispersion to grow. For strata 4, 5, and 6, those show less variability in the housing price per square meter than housing prices in stratum 2.
\begin{figure}[H]
    \centering
    \includegraphics[scale=0.65]{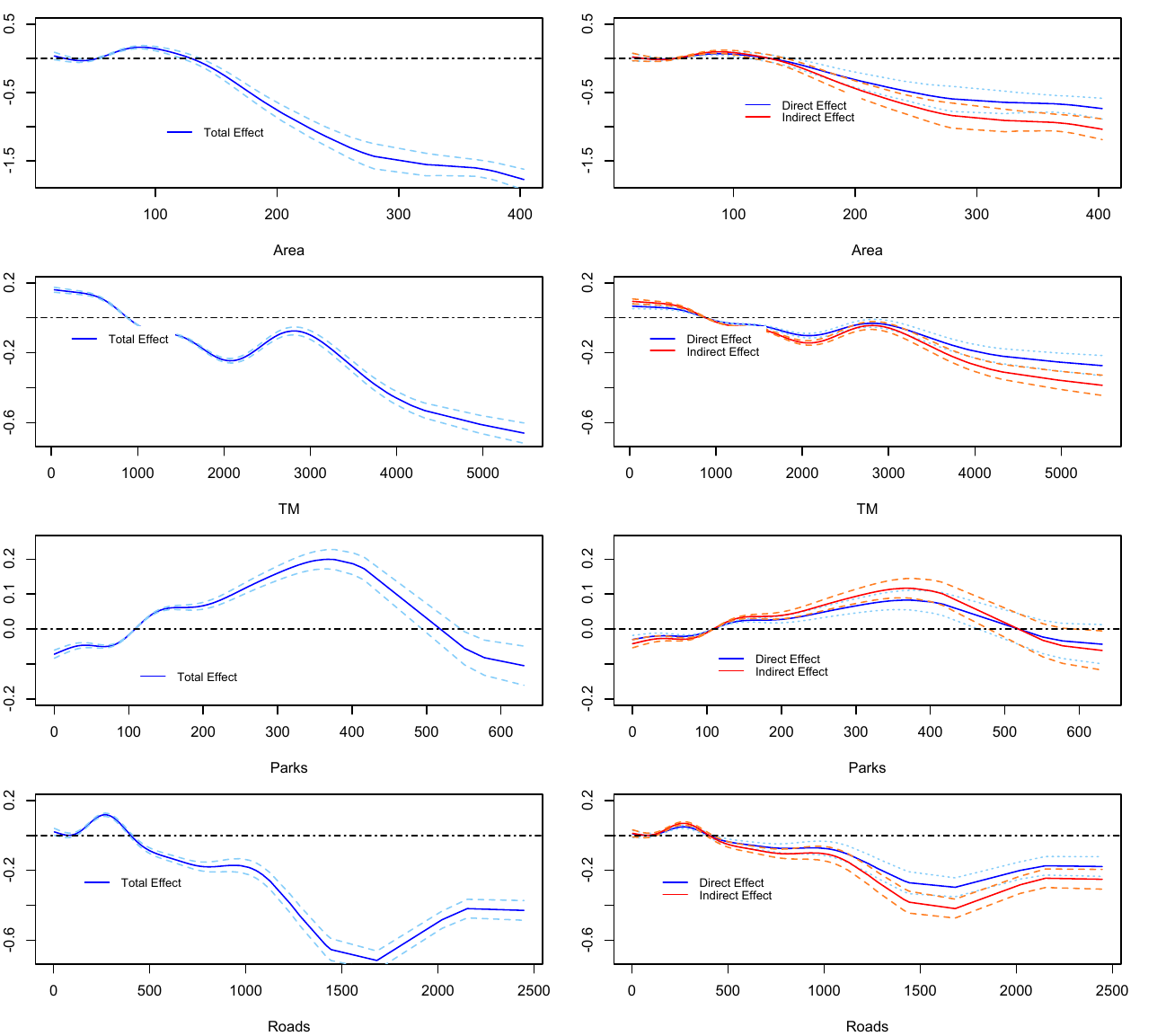}
        \caption{Estimated effects over the mean non-parametric functions for the variables included in the hedonic price model for Bogotá city with semiparametric SAR with non-homogeneous variance.}
        \label{nueva_bogota}
\end{figure}

\begin{figure}[H]
    \centering
    \includegraphics[scale=0.45]{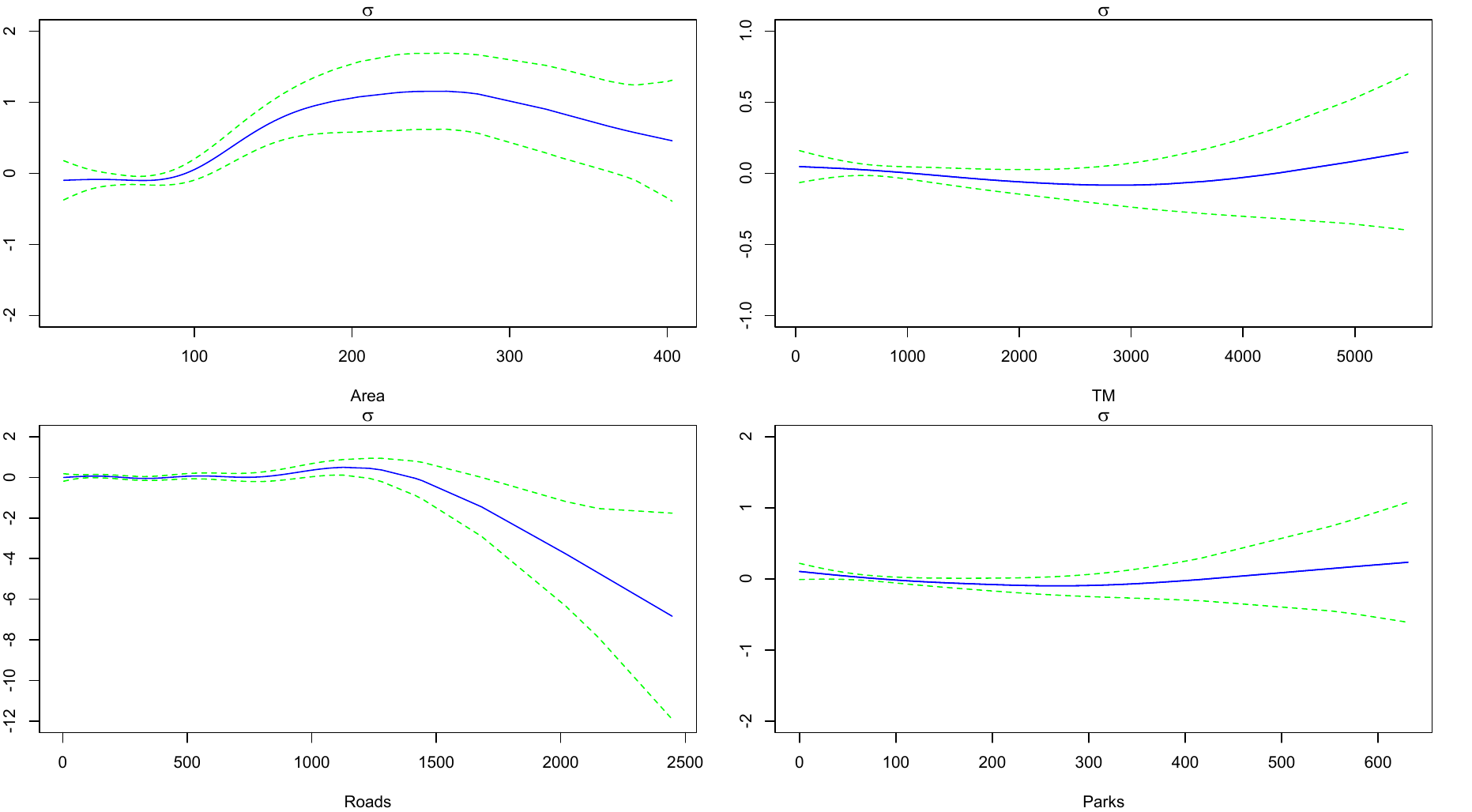}
        \caption{Estimated effects over the variance non-parametric functions for the variables included in the hedonic price model for Bogotá city with semiparametric SAR with non-homogeneous variance.}
        \label{nueva_bogota1}
\end{figure}

On the other hand, \textcolor{red}{the estimated smooth functions cannot be interpreted as elasticities. Taking advantage of the results obtained by \cite{basile2014modeling}, the direct, total, and indirect effects for each non-parametric term can be calculated. When analyzing Figure \ref{nueva_bogota}, the existence of non-linear relationships between some of these variables and the logarithm of the price of a property is evident, where the flexibility of the GAMLSS models is better used in the estimation of the standard deviation (see Figure \ref{nueva_bogota1}). In this way, there is an increasing relationship between the house size and the price per square meter, starting from 80 square meters because before this relationship was negative. The price per square meter will be lower, the farther away a home is from a Transmilenio station.}

\textcolor{red}{A decreasing marginal relationship up to 400 meters is found in the distance to parks; from this point on, the housing price begins to reduce.} This situation is explained by the perception of insecurity generated by neighborhood and small-scale parks. The distance to main roads has a clear non-linear behavior for the price per square meter, which would not have been captured if other models such as traditional SAR had been used.

\begin{figure}[H]
     \centering
     \includegraphics[scale=0.43]{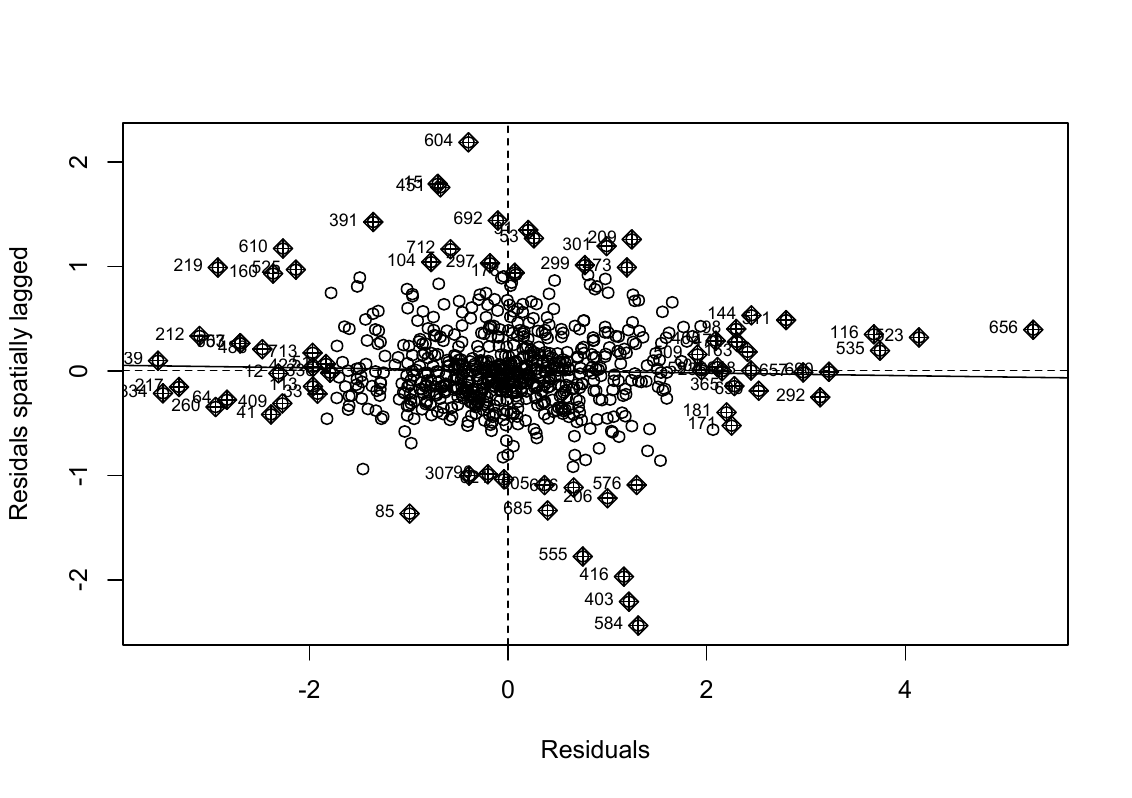}
     \caption{Moran's I plot on the residuals obtained from the model built for the hedonic prices of Bogotá City with semiparametric SAR and non-homogeneous variance.}
     \label{moran_final}
\end{figure}

After 100 square meters, there is an increase in the variability of the price per square meter. However, the variability is reduced in properties that exceed 250 square meters. The distance to Transmilenio shows a small u-shaped relationship with the variability of the price per square meter. The variability of the price per square meter is constant for the distance to main roads. After 1200 meters, the price per square meter of the properties will tend to be very similar to each other (lower variance). The distance to parks has a slightly increasing relationship concerning the variability of the price per square meter of real estate, this is evident after 300 meters. Finally, Figure \ref{moran_final} shows Moran's I graph for the residuals generated from the proposed methodology, where the absence of spatial dependence is clear, which shows the usefulness of the method to treat this type of situation.

\section{Conclusions}

Semiparametric SAR models with non-homogeneous variance allow greater flexibility and the ability to adapt to the different situations that arise when working with spatial data. In this sense, the model obtains two properties that give it greater advantages over other linear and non-parametric methodologies: i) to allow the modeling of the spatial autoregressive parameter ($\rho$) appropriately and ii) to estimate linear and non-linear relationships for both the mean and the variance.
The above is conceived as an important gain because the traditional GAMLSS of \cite{rigby2005generalized} models, the mean and variance \textcolor{red}{includes non-parametric components, but spatial modeling is carried out through GMRF \citep{de2018gaussian} and not from the traditional approach formulated by \citet{anselin1988spatial}}. 
On the other hand, the proposal of \cite{montero2012sar} only estimates the parameters associated with the mean, assuming that the variance is constant over all observations.
The method to estimate our proposed model was obtained by adapting the traditional spatial regression estimation algorithms and those used in GAMLSS. Furthermore, when carrying out the simulations, a great recovery capacity of the parameters was evident under different scenarios of sample size and values for $\rho$.

The proposed methodology was applied effectively in the estimation of a hedonic price model for Bogotá city, finding interesting results by allowing the incorporation of non-linear relationships between the explanatory variables and the dependent one (both for the mean and the standard deviation), with an estimation of the spatial autoregressive parameter that was significant. This exercise is conceived as an extension of the work developed by \cite{delgado2021determinants}.
On the other hand, this algorithm can be used for completely linear models, where the variance or standard deviation can be modeled, its performance is comparable to the Bayesian methodologies proposed by \cite{sicachapackage}.
Finally, it is important to highlight that although this work was oriented towards analysis in real estate markets, these models can be easily applied to situations and analyses associated with agriculture and ecology \citep{plant2018spatial,ver2018spatial}.
The proposed methodologies are compatible with the estimates made through the mgcv \citep{wood2015package} and GAMLSS 
 \citep{stasinopoulos2020package} packages of the \citet{Rmanual}, where the latter allows modeling the scale parameter as variance ($\sigma^2$) or as standard deviation ($\sigma$).
Likewise, the step to follow will be to include time within the estimate, thus allowing it to work with data from repeated measures or panels (spatio-temporal models) since only cross-sectional data were considered.

Finally, it would be valuable to extend these models to other hierarchical or multilevel structures such as those proposed by \cite{cellmer2019application}, where in the case of properties, two neighborhood structures are considered: the first between properties (disaggregated) and the second from the areas where said homes are located (aggregated). The above could allow a better evaluation of interactions and control of spatial heterogeneity.

\section*{Supplementary files}
\begin{enumerate}
    \item Supplementary file 1\label{sf1}: File with the simulations obtained for the regular grid, irregular grid, and spatial points of each of the compared models.
    \item Supplementary file 2\label{sf2}: R files with the three simulations of this paper.
\end{enumerate}

\appendix
\section{Analysis on an irregular grid} \label{simCundi}

\begin{figure}[H]
\centering
\includegraphics[scale=0.35]{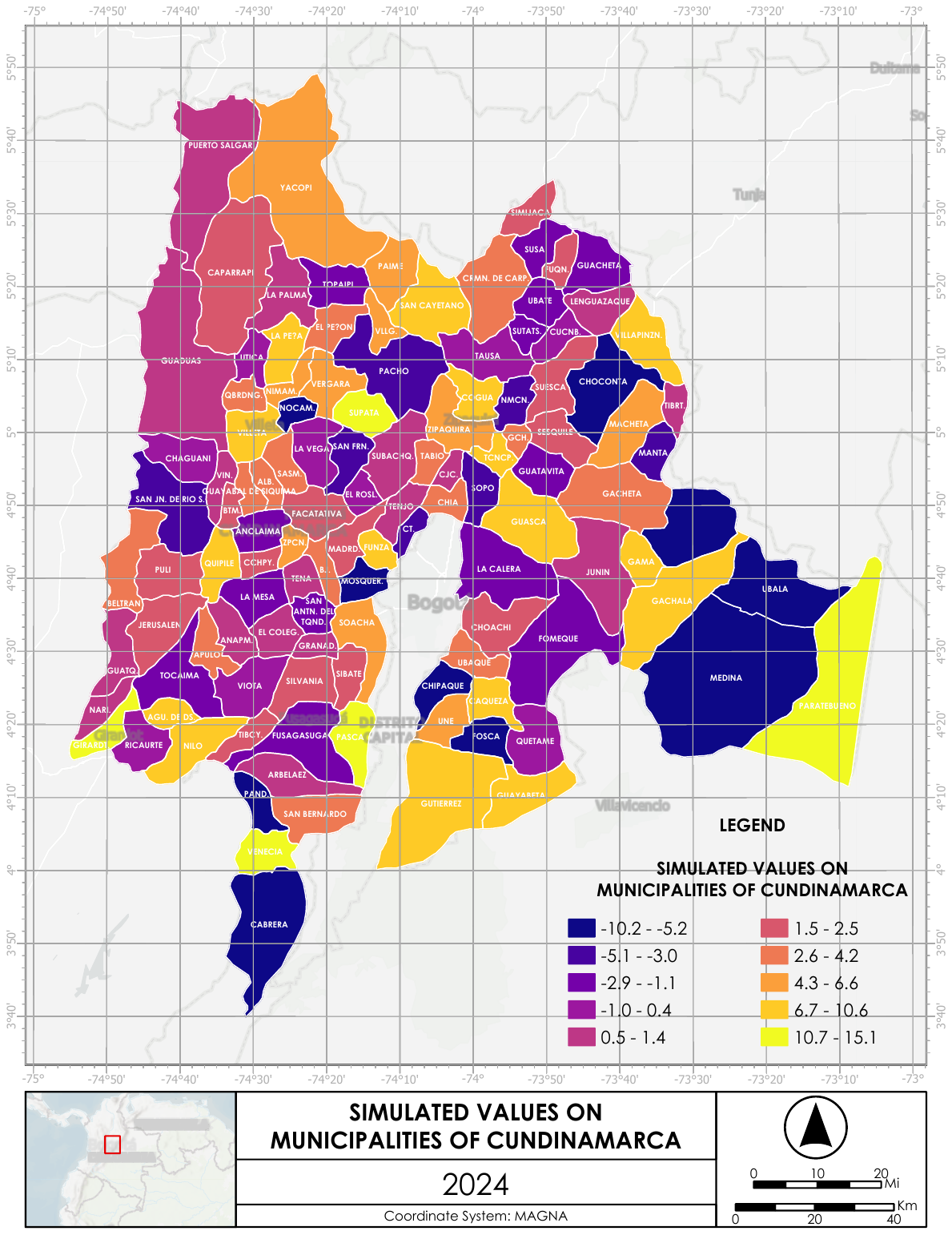}
\caption{Example of simulated values on municipalities of Cundinamarca (Colombia)}
\label{mapa_cundi}
\end{figure}

After analyzing the results of the simulations on a regular grid, a similar procedure will be carried out for the set of 116 municipalities belonging to the department of Cundinamarca (Colombia). For this exercise, the explanatory variables were generated from a uniform distribution $x_{1i}\sim U(1,10)$, $x_{2i}\sim U(0,1)$ and $x_{3i}\sim U(0,1)$ and different values were set for the parameter $\rho =\{-0.75, -0.4, 0.4, 0.75\}$, \textcolor{red}{The other parameters of the simulation were left with the same specification as the regular grid.} Figure \ref{mapa_cundi} shows the result of one of these simulations on the map.

\begin{figure}[H]
\centering
\includegraphics[scale=0.57]{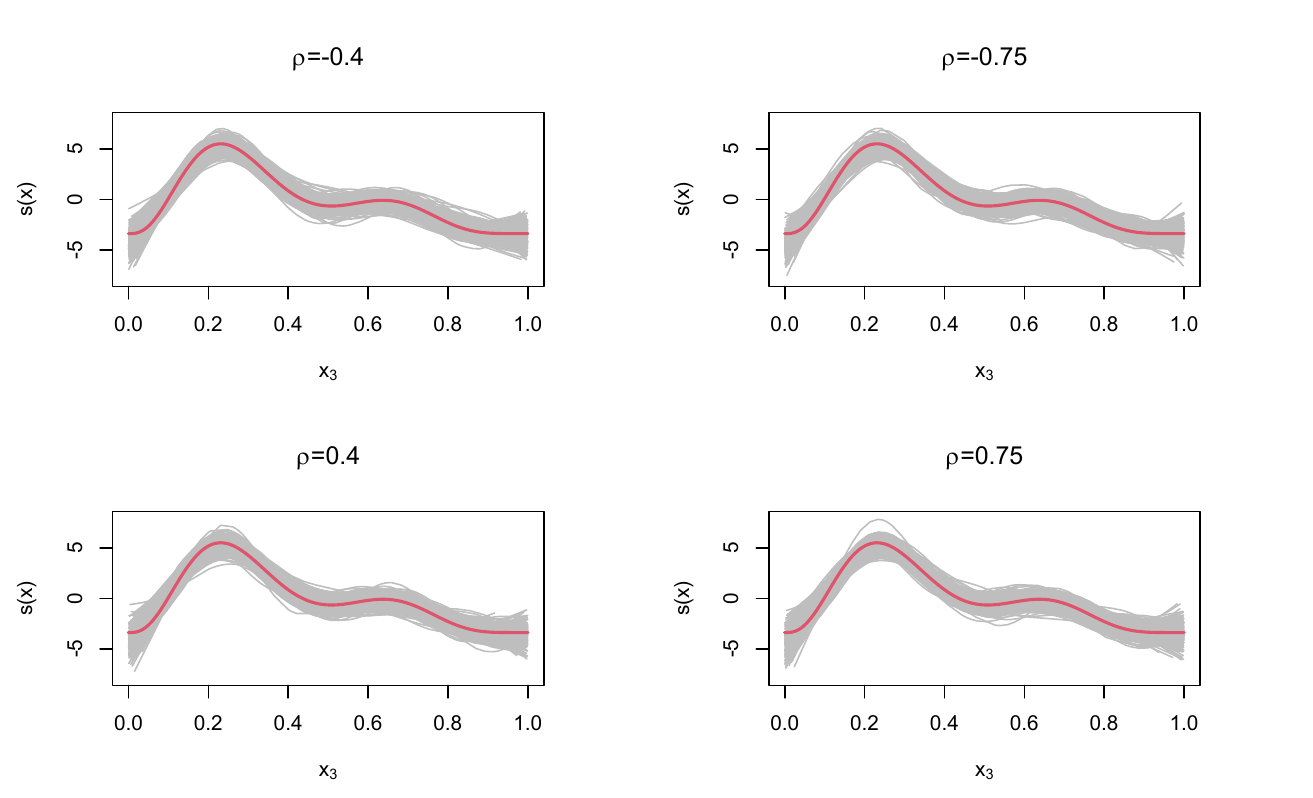}
\caption{Estimated functions under different values of $\rho \in \{-0.75, -0.4, 0.4, 0.75\}$ for the simulation on an irregular grid.}
\label{curvas_cundi}
\end{figure}

Taking into account that the sample size is fixed (116 municipalities), the simulations will focus on evaluating the performance of the algorithm under different values of the spatial autoregressive parameter.
The results are similar to those found under a perfect grid, showing a high recovery capacity of the original parameters for both the mean and the standard deviation. Likewise, this behavior is maintained for different values of $\rho$ (see Table \ref{cundi_tabla}).
On the other hand, when analyzing Figure \ref{curvas_cundi}, the estimated curves have the same behavior as the original function, which shows the capacity of the model to adjust this type of non-parametric component.

\begin{table}[H]
    \centering
    \begin{tabular}{cc}
    \begin{tabular}{|c|c|c|}
    \hline
      $\rho$ & Par & Estimates\\
       & & \begin{tabular}{ccc}
            Mean & Sd & Bias\\
       \end{tabular}\\
       \hline
      0.4 & \begin{tabular}{c}
           $\hat{\rho}$\\
           $\hat{\beta}_0$\\
           $\hat{\beta}_1$\\
           $\hat{\beta}_2$\\
           $\hat{\alpha}_0$\\
            $\hat{\alpha}_0$
      \end{tabular}& \begin{tabular}{ccc}
         0.384 & 0.081 & -0.040  \\ 
 5.471 & 0.743 & 1.735  \\ 
-0.494 & 0.075 & -0.011 \\ 
 1.734 & 0.625 & -0.009  \\ 
 0.417 & 0.160 & -0.166\\ 
   0.343 & 0.271 & 0.144  
      \end{tabular}\\
            \hline
             0.75 & \begin{tabular}{c}
           $\hat{\rho}$\\
           $\hat{\beta}_0$\\
           $\hat{\beta}_1$\\
           $\hat{\beta}_2$\\
           $\hat{\alpha}_0$\\
            $\hat{\alpha}_0$
      \end{tabular}& \begin{tabular}{ccc}
  0.734 & 0.051 & -0.022  \\ 
 5.640 & 0.965 & 1.820  \\ 
-0.498 & 0.075 & -0.003  \\ 
 1.727 & 0.673 & -0.012  \\ 
0.416 & 0.178 & -0.168  \\ 
 0.341 & 0.305 & 0.138 
      \end{tabular}\\
        \hline
    \end{tabular}
& 
    \begin{tabular}{|c|c|c|}
    \hline
      $\rho$ & Par & Estimates\\
       & & \begin{tabular}{ccc}
            Mean & Sd & Bias\\
       \end{tabular}\\
       \hline
     -0.4 & \begin{tabular}{c}
$\hat{\rho}$\\
$\hat{\beta}_0$\\
$\hat{\beta}_1$\\
$\hat{\beta}_2$\\
$\hat{\alpha}_0$\\
$\hat{\alpha}_0$
  \end{tabular}& \begin{tabular}{ccc}
-0.406 & 0.104 & 0.016  \\ 
5.388 & 0.611 & 1.693  \\ 
-0.497 & 0.068 & -0.005  \\ 
1.800 & 0.639 & 0.028  \\ 
0.420 & 0.164 & -0.159 \\ 
0.332 & 0.288 & 0.105  \\ 
\end{tabular}\\
\hline
-0.75 & \begin{tabular}{c}
$\hat{\rho}$\\
$\hat{\beta}_0$\\
$\hat{\beta}_1$\\
$\hat{\beta}_2$\\
$\hat{\alpha}_0$\\
$\hat{\alpha}_0$
  \end{tabular}& \begin{tabular}{ccc}
-0.747 & 0.094 & -0.003 \\ 
5.375 & 0.624 & 1.687 \\ 
-0.496 & 0.073 & -0.008 \\ 
1.781 & 0.676 & 0.017 \\ 
0.419 & 0.162 & -0.162 \\ 
0.344 & 0.277 & 0.146 \\
\end{tabular}\\
\hline
    \end{tabular}
    \end{tabular}

    \caption{Results of the simulations of the parametric effects over mean and standard deviation models with $\beta_0=2$, $\beta_1$=-0.5, $\beta_2$=1.75, $\alpha_0$=0.5 and $\alpha_1$=0.3 on an irregular grid.}
    \label{cundi_tabla}
\end{table}
\textcolor{red}{
In the supplementary file \ref{sf1}, it is evident that the proposed estimation model has similar results to those obtained from the simulations carried out on a regular grid and with tower-like contiguity. Consequently, the conclusions previously found are maintained under this scenario and prove that H-AM-SAR obtains good performance in different situations and improves the models that were adjusted.  So, the proposed method can be extended to completely parametric models, becoming an estimation alternative that is not computationally intensive.}
\end{document}